\documentclass{aa}
\usepackage[varg]{txfonts}
\usepackage{booktabs,graphicx,siunitx,xspace,natbib}
\usepackage{subfig}
\usepackage{ragged2e} 
\usepackage{array}
\usepackage{multirow}
\DeclareSIUnit \parsec {pc}
\DeclareSIUnit \pc     {\parsec}
\DeclareSIUnit \kpc    {\kilo \parsec}
\DeclareSIUnit \msun   {\ensuremath{M_\sun}}
\DeclareSIUnit \year   {a}

\newcommand{\erfc}{\mathrm{erfc}}
\newcommand{\Msol}{M\ensuremath{_\odot}\xspace}

\begin{document}

\title{Gamma-ray spectroscopy of positron annihilation in the Milky Way}

\author{
  Thomas Siegert   		   \inst{\ref{inst:mpe}}\thanks{E-mail: tsiegert@mpe.mpg.de} \and
  Roland Diehl           \inst{\ref{inst:mpe},\ref{inst:xcu}} \and
  Gerasim Khachatryan    \inst{\ref{inst:mpe}} \and
  Martin G. H. Krause    \inst{\ref{inst:usm},\ref{inst:xcu},\ref{inst:mpe}} \and	
  Fabrizia Guglielmetti  \inst{\ref{inst:mpe}} \and	
  Jochen Greiner         \inst{\ref{inst:mpe},\ref{inst:xcu}} \and
  Andrew W. Strong       \inst{\ref{inst:mpe}}     \and
  Xiaoling Zhang         \inst{\ref{inst:mpe}}    % \and
  %{\it et al.}
}
\institute{
  Max-Planck-Institut f\"ur extraterrestrische Physik, Gie\ss enbachstr. 1, D-85741 Garching, Germany
  \label{inst:mpe}
  \and
  Excellence Cluster Universe, Boltzmannstr. 2, D-85748, Garching, Germany
  \label{inst:xcu}
  \and
  Universit\"ats-Sternwarte Ludwig-Maximilians-Universit\"at,
  D-81679 M\"unchen, Germany
  \label{inst:usm}
  }
\date{Received 12 Oct 2015 / Accepted 30 Nov 2015}

\abstract
% context heading (can be left empty)
{The annihilation of positrons in the Galaxy's interstellar medium produces characteristic gamma-rays with a line at 511~keV. This gamma-ray emission has been observed with the spectrometer SPI on ESA's INTEGRAL observatory, confirming a puzzling morphology with bright emission from an extended bulge-like region, while emission from the disk is faint. Most known or plausible sources of positrons are, however, believed to be distributed throughout the disk of the Milky Way.}
% aims heading
{We aim to constrain characteristic spectral shapes for different spatial components in the disk and bulge using data with an exposure that has doubled since earlier reports.}
% methods heading
{We exploit high-resolution gamma-ray spectroscopy with SPI on INTEGRAL based on a new instrumental background method and detailed multi-component sky model fitting.}
% results heading
{We confirm the detection of the main extended components of characteristic annihilation gamma-ray signatures, altogether at 58$\sigma$ significance in the 511~keV line. The total Galactic 511~keV line intensity amounts to $(2.74\pm0.25)\times10^{-3}~\mathrm{ph~cm^{-2}~s^{-1}}$ for our assumed model of the spatial distribution. We derive spectra for the bulge and disk, and a central source modelled as point-like and at the position of Sgr A*, and discuss spectral differences. The bulge ($56\sigma$) shows a 511~keV line intensity of $(0.96\pm0.07)\times10^{-3}~\mathrm{ph~cm^{-2}~s^{-1}}$ together with ortho-positronium continuum equivalent to a positronium fraction of ($1.080\pm0.029$). The two-dimensional Gaussian that represents the disk emission ($12\sigma$) has an extent of $60^{+10}_{-5}$~degrees in longitude and a rather large latitudinal extent of $10.5^{+2.5}_{-1.5}$~degrees; the line intensity is $(1.66\pm0.35)\times10^{-3}~\mathrm{ph~cm^{-2}~s^{-1}}$ with a marginal detection of the annihilation continuum and an overall diffuse Galactic continuum of $(5.85\pm1.05)\times10^{-5}~\mathrm{ph~cm^{-2}~s^{-1}~keV^{-1}}$ at 511~keV. The disk shows no flux asymmetry between positive and negative longitudes, although spectral details differ. The flux ratio between bulge and disk is ($0.58\pm0.13$). The central source ($5\sigma$) has an intensity of $(0.80\pm0.19)\times10^{-4}~\mathrm{ph~cm^{-2}~s^{-1}}$.}
% conclusions heading (can be left empty)
{}


\maketitle

% ----------------------------------------------------------------------
%
\section{Introduction}
\label{sec:intro}

Positrons in our galaxy have been studied with many different astronomical instruments through the gamma-rays at the characteristic energy of 511 keV produced during their annihilation in the interstellar medium (ISM). Following early reports of a line near this energy, the first high resolution measurements~\citep{Leventhal1978_511} established its identification as positron annihilation. Early indications of variability were later shown to have been due to comparing measurements of extended emission made with instruments having different fields of view~\citep{Albernhe1981_511,Leventhal1986_511,Share1988_511}.

Two remarkable transient annihilation events found with SIGMA \citep[e.g.][]{Bouchet1991_mq511,Goldwurm1992_511} were apparently associated with a hard X-ray source, 1E1740.7-2942, near the centre of the galaxy, that was consequently termed the ``Great Annihilator''. Neither contemporaneous~\citep{Jung1995_mq511,Smith1996_mq511crab} nor long-term monitoring of the source  \citep{Sunyaev1991_mq,Harris1994_transient511,Smith1996_transient511,Cheng1998_transient511} provided confirmation of such activity.

OSSE on the Compton Gamma-Ray Observatory (CGRO) made extensive observations of the 511 keV emission, confirming the nature of the emission as constant, extended along the galactic plane with a strong concentration towards the Galactic centre \citep{Purcell1993_511,Purcell1997_511}. Deconvolved sky maps appeared to show an extension of the emission in the galactic centre region towards positive latitudes~\citep{Dermer1997_511}. 

Prior to the launch in 2002 of the INTEGRAL spacecraft~\citep{Winkler2003_INTEGRAL}, information had been obtained with little imaging capability and often with limited spectral resolution. For example, the field of view (FoV) of the OSSE collimators was 3.8 by 11 degrees and its spectral resolution at 511 keV about 10\%. Observations with the good energy resolution of Germanium detectors had been obtained only with wide field instruments on balloons and on HEAO-3. The SPI gamma-ray spectrometer instrument on INTEGRAL provided a significant advance, with a coded mask allowing imaging at $\sim$2 degree precision, and Ge detectors with $\sim$2~keV intrinsic spectral resolution~\citep{Vedrenne2003_SPI,Roques2003_SPI}.

The INTEGRAL sky survey has excellent exposure of the entire inner Galaxy. This led to a first all-sky image of positron annihilation gamma-rays~\citep{Knoedlseder2003_511,Knoedlseder2005_511}. INTEGRAL data confirmed that annihilation gamma-ray emission is dominated by a bright and extended region centred in the Galaxy; only relatively weak emission was seen from the extended plane of the Milky Way outside the central region. The morphology was found to be much more symmetric about the Galaxy's centre and to have an extent of 10--12$^\circ$ (FWHM); no extension towards northern latitudes was seen. Attempts to separate bright bulge from faint disk emission led to some discussion and confusion about an ``asymmetry'' \citep{Weidenspointner2008a_511,Higdon2009_511,Skinner2010_511}. But later, as the disk emission was found, the apparent asymmetry was more readily explained by a slight offset ($\sim1^\circ$) of the centroid of the bright bulge-like model \citep{Bouchet2010_511,Skinner2012_511}. 

The spectrum of positron annihilation measured with SPI demonstrated that the dominating line at 511 keV was centred at the expected energy, and slightly broadened \citep{Jean2006_511}. The presence of the characteristic continuum from three-photon annihilation, already pointed out from earlier measurements \citep[e.g.][]{Leventhal1978_511,Kinzer2001_511}, was also confirmed. The interpretation of positron annihilation occurring in a partially ionised medium with typical temperatures near 8000~K was supported~\citep{Churazov2005_511,Churazov2011_511}. 

Candidate positron sources have been reviewed by \citet{Prantzos2011_511}, and include:
\begin{enumerate}
	\item Radioactive decay of $\beta^+$ unstable isotopes, such as $^{56}$Ni, $^{44}$Ti, $^{26}$Al, or $^{13}$N, produced in nucleosynthesis sources throughout the Galaxy;
	\item Accreting binary systems, producing jets loaded with pair plasma, microquasars being the prominent examples;
	\item Pulsars, because curvature radiation produces and ejects pair plasma;
	\item The supermassive black hole in our Galaxy's centre (Sgr A*) through various mechanisms;
	\item Dark matter decay or annihilation, as dark matter would be gravitationally concentrated in the inner Galaxy.
\end{enumerate}

Although each of these sources have been studied in some detail~\citep[e.g.][for nucleosynthesis, microquasars, and dark matter, respectively]{Martin2012_511,Guessoum2006_MQ511,Boehm2004_dm}, quantitative estimates of their contributions leave considerable uncertainties~\citep[see discussions in][and their Table IX, see also~Sect.~\ref{sec:discussion} of this paper]{Prantzos2011_511}. The locations and distributions of the various possible sources within the Galaxy offer a way to constrain their relative importance, if annihilation is assumed to occur close to the respective sources ($\lesssim$ 100~pc)~\citep{Prantzos2011_511}. Most of the above candidate sources would be distributed along the disk of the Milky Way; Sgr A* and dark matter contributions would be concentrated in the central region of the Galaxy, and some sources such as supernovae of type Ia (SNe Ia) or low-mass X-ray binaries (LMXRBs) might plausibly be related to an extended bulge or a thick disk. But combined with expected source numbers and positron yields, the puzzle still remains, and no conclusive candidate has emerged.

More complexities arise because once ejected from their sources, the positron annihilation may occur at a distance far from their origins. Because the positrons are produced at high energy (MeV to GeV), many positrons will leave the source, and propagate in the surrounding interstellar gas. Positrons slow down to energies of a few eV and annihilation may occur. The propagation of positrons in the Galaxy has been extensively investigated~\citep{Guessoum1991_511ISM,Guessoum2005_511,Jean2009_511ISM,Alexis2014_511ISM}. By the detection of cosmic ray electrons at lower energies ($\lesssim\mathrm{MeV}$), it has been shown that they can propagate on kpc scales in the ISM~\citep[e.g.][]{Lingenfelter2009_dm511}. Detailed propagation calculations~\citep{Higdon2009_511,Alexis2014_511ISM} have shown that positrons may propagate up to kpc distances or more when they face a very low density, and hot ISM phase before annihilation. In contrast, when positrons enter the much denser, warm cloud phases, i.e. the region where the annihilation takes place~\citep{Jean2006_511,Churazov2005_511,Churazov2011_511}, the propagation distances are smaller. Thus, irrespective of whether the positrons are produced in young massive stars near the clouds, or, more diffusely, on kpc scales in SNe Ia of much older, accreting white dwarves, they will annihilate predominantly in these warm phases. The distribution of such gas, weighted by source and propagation effects, is what is essentially observed in the 511~keV line emission, and not directly their source distribution.

% ----------------------------------------------------------------------
\begin{figure}
  \centering
  \includegraphics[width=\linewidth]{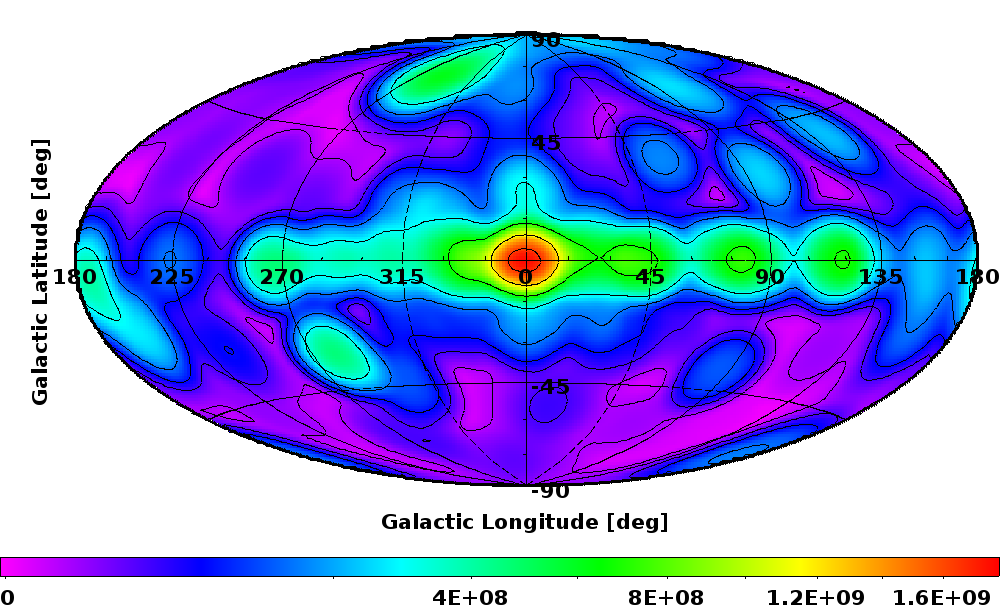}
   \caption{Sky exposure with SPI for the data set analysed. The units of the map are given in $\mathrm{cm^2 s}$. The equivalent exposure time is calculated for 19 detectors with an effective area of $\sim75~\mathrm{cm^2}$ for photon energies around 511 keV. The contours, starting from the innermost, correspond to exposures of 22~Ms, 16~Ms, 9~Ms, 4~Ms, 2~Ms, 1~Ms, 0.5~Ms, and 0.1~Ms, respectively. The total exposure time is 160~Ms.}
  \label{fig:SPI-exposure}
\end{figure}
% ----------------------------------------------------------------------

In this paper, we report a new study of measurements with INTEGRAL/SPI data accumulated over eleven years. We employ coded mask imaging together with high-resolution spectroscopy, enhanced by a new approach to instrumental background spectra. We aim at determining details of the annihilation spectra, in particular the 511~keV line centroid energy and broadening, which characterises the kinematics and temperature of the annihilation region, and also the annihilation line-to-continuum ratio, which characterises the fraction of annihilations that take place via the formation of a positronium atom. Our analysis discriminates between the spectra of the bulge and disk regions, and of candidate point sources.    

% ----------------------------------------------------------------------
\begin{figure}
  \centering
  \includegraphics[trim=3cm 5cm 3cm 5cm,clip,width=\linewidth]{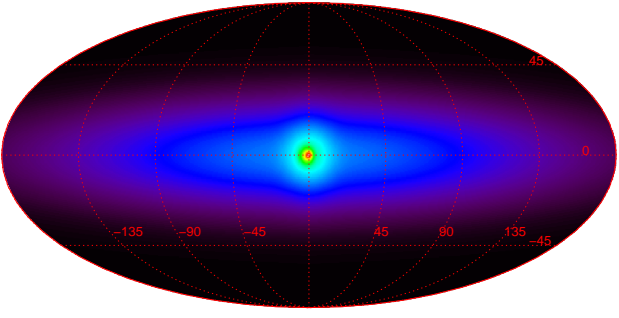}
    \caption{Image showing the model components as assessed in the sky model fit study. The components are related to the ones from \citet{Skinner2014_511}, see Tab.~\ref{tab:model_components} for details. The weighted sum of fluxes for each celestial component in the 80 bins in the analysed energy band from 490 to 530~keV is shown. Two additional point sources to improve the fit, the Crab and Cyg X-1, are not shown. The image has been scaled by taking the cube root to emphasise the low surface-brightness and extent of the disk.}
  \label{fig:SPI-imagemodel}
\end{figure}
% ----------------------------------------------------------------------

% ----------------------------------------------------------------------
%
\section{Data and their Analysis}
\label{sec:spi-analysis}
% ----------------------------------------------------------------------

\subsection{Instrument and data}

The SPI camera utilises the \emph{coded mask} technique for imaging gamma-ray sources. In this technique, a mask with occulting tungsten blocks and holes in the aperture of the gamma-ray camera imprints a shadowgram of a celestial source onto a multi-element detector array. This array consists of 19 high-resolution Ge detectors that measure photons between 20~keV and 8~MeV with a spectral resolution of $\sim2.2$~keV at 662~keV. Data consist of energy-binned spectra for each pointing on the sky and each of the 19 Ge detectors. During the ongoing INTEGRAL mission, four of the SPI detectors failed, reducing the effective area.

Exposures are taken in successive pointings of the telescope of typically \SI{1800}{\second} duration, moving the telescope pointing by $\sim$2 degrees along a rectangular $5\times5$ pattern between pointings to shift the shadowgram of the source in the detector plane. Surveys concatenate adjacent sky region pointing patterns; the FoV of SPI is $16^{\circ}\times16^{\circ}$ (partially coded FoV: $34^{\circ}\times34^{\circ}$). Instrumental background practically remains constant between adjacent pointings, thus allowing it to be discriminated against the mask-encoded signal from the sky.

For our analysis, we used exposures of various parts of the Galaxy, accumulated over eleven years of the INTEGRAL mission, during orbit numbers 21 to 1279, with gaps due to enhanced solar activity, calibration, or annealing periods. The exposures for particular regions in the sky are given in Tab.~\ref{tab:expo}. 
% ----------------------------------------------------------------------
\begin{table}[!ht]
\caption{Exposure times in the regions of the sky used in this work. The times have been extracted from the exposure map (Fig.~\ref{fig:SPI-exposure}) in the listed longitude ranges and for latitudes $|b|<45^{\circ}$ in each case. $T^+$ and $T^-$ are the exposure times for positive and negative longitudes, respectively, given in Ms.}              % title of Table
\label{tab:expo}      % is used to refer this table in the text
\centering                                      % used for centering table
%\begin{tabular}{>{\RaggedRight}p{1.0cm} >{\RaggedLeft}p{1.4cm} >{\RaggedLeft}p{1.4cm} >{\RaggedLeft}p{1.6cm} >{\RaggedLeft}p{1.6cm}}          % centered columns (6 columns)
\begin{tabular}{c | c c c c}
\hline\hline                        % inserts double horizontal lines
                                   % inserts single horizontal line
    $|l|$    & $<15^{\circ}$   & $15^{\circ}-60^{\circ}$ & $60^{\circ}-105^{\circ}$   & $105^{\circ}-180^{\circ}$  \\      % inserting body of the table
    $T^+$    & \multirow{2}{*}{$30.8$}          & $21.8$                   & $17.6$                      & \multirow{2}{*}{$30.7$} \\
    $T^-$    &                                  & $20.4$                   & $15.6$                       \\
%    Regions  & Inner Galaxy   & Inner Disk & Cygnus / Carina  & Orion / Crab  \\
\hline                                             %inserts single line
\end{tabular}
\end{table}
% ----------------------------------------------------------------------

The difference in exposure between positive and negative longitudes, integrated over all latitudes, is less than 7\% (see Fig.~\ref{fig:SPI-exposure}). Data selections are applied to suppress contamination, e.g. from solar flare events. Our dataset thus consists of 160~Ms of exposure, with 73590 telescope pointings. Taking into account the detectors failures, 1214799 individual spectra are to be analysed.

\subsection{Analysis method}

Our analysis method is based on a comparison of measured data with predicted data from models. The comparison is performed in a data space consisting of the counts per energy bin measured in each of SPI's detectors for each individual exposure (pointing) as part of the complete observation.

We describe data $d_k$ per energy bin $k$ as a linear combination of the sky contribution, i.e. model
components $M_{ij}$, to which the instrument response matrix $R_{jk}$
is applied for each image element $j$, and the background, i.e. components $B_{ik}$ for line and continuum instrumental background. Scaling parameters $\theta_i$ for $N_I$ are provided for sky and $N_B$ background components:
\begin{equation}
d_k = \sum_j R_{jk} \sum_{i = 1}^{N_\mathrm{I}} \theta_i M_{ij} + \sum_{t}\sum_{i = N_\mathrm{I} + 1}^{N_\mathrm{I} + N_\mathrm{B}} \theta_{i,t} B_{ik}\mathrm{.}
\label{eq:model-fit}
\end{equation}
We fit these scaling parameters, using the maximum likelihood technique, applied to energy bins covering the spectral range of interest. The energy band used, from 490 to 530~keV with 0.5~keV energy bins, is chosen to allow the study of the shape of the 511~keV annihilation line, together with the ortho-positronium continuum, and a Galactic gamma-ray continuum. The scaling parameters $\theta_i$ for the $N_I$ sky components are set constant in time, while the scaling parameters for the $N_B$ background components, $\theta_{i,t}$, are allowed to vary with time $t$ (see section~\ref{sec:bg_model}).

For each camera configuration, corresponding to a given number of working detectors, a specific imaging response function is applied to each of the sky model components to account for the shadowgram of the mask. These response functions are different for the off-diagonal terms, which account for scattering in dead detectors followed by detection in another detector. This effect creates a tail in the expected spectrum towards lower energies. For photons between 490 and 511~keV, this tail contains about 3\% of the line flux~\citep[see also][]{Churazov2011_511}.

\subsection{Celestial emission modelling}
\label{sec:sky_emission}

In our spectral fits, we use a multi-component description of the distribution of the emission over the sky. In a recent analysis of a similar data set in a single 6~keV energy bin, centred on the 511~keV line, \citet{Skinner2014_511} propose a representation of the positron annihilation sky in which the emission from the disk is represented by a two-dimensional Gaussian function with different widths in longitude and latitude, and that from the bulge as the sum of three components: two symmetrical three-dimensional Gaussians and a third component which is consistent with a point source. One of the Gaussians representing the bulge is offset to negative longitudes while the other components are centred at the Galactic centre (in the case of the point-like component, the position is actually taken as that of Sgr A*, see Fig.~\ref{fig:SPI-imagemodel}). Point sources are added at the positions of the Crab, and Cyg X-1. Although such modelling includes correlations among components, it can be seen as an alternative to having a large number of pixels on the sky or orthogonalised functions that have no astrophysical basis, as it associates sky components with plausible and explicit source regions. The six components used for modelling the celestial emission in the energy range from 490 to 530~keV are listed in Tab.~\ref{tab:model_components}. As the latitude and longitude extent of the disk are considered the most uncertain parameters, for our spectroscopic analysis in fine energy bins, we scan the plausible parameter region with 100 different disk shapes/extents in both, longitude and latitude width (see Sect.~\ref{sec:disk}).

% ----------------------------------------------------------------------
\begin{table}
\caption{Characteristics of the sky model components assumed in our analysis. The parameters are similar to those of \citet{Skinner2014_511}, except for the extent of the disk\tablefootmark{a}.}              % title of Table
\label{tab:model_components}      % is used to refer this table in the text
\centering                                      % used for centering table
\begin{tabular}{>{\RaggedRight}p{1.45cm} >{\RaggedLeft}p{1.3cm} >{\RaggedLeft}p{1.3cm} >{\RaggedLeft}p{1.6cm} >{\RaggedLeft}p{1.55cm}}          % centered columns (6 columns)
\hline\hline                        % inserts double horizontal lines
%Name & Galactic Longitude Position [deg] & Galactic Latitude Position [deg] & Longitude extent (FWHM) [deg] & Latitude extent (FWHM) [deg] \\
Comp. & G. Lon. position [deg] & G. Lat. position [deg] & Lon. extent (FWHM) [deg] & Lat. extent (FWHM) [deg] \\
\hline                                   % inserts single horizontal line
    NB                        & $-1.25$   & $-0.25$ & $5.75$   & $5.75$  \\      % inserting body of the table
    BB                        & $0.00$    & $0.00$  & $20.55$  & $20.55$ \\
    Disk\tablefootmark{a}     & $0.00$    & $0.00$  & $141.29$ & $24.73$ \\
    GCS\tablefootmark{b}\tablefootmark{c}      & $-0.06$   & $-0.05$ & $0.00$   & $0.00$  \\
    Crab\tablefootmark{b}     & $-175.44$ & $-5.78$ & $0.00$   & $0.00$  \\
    Cyg X-1\tablefootmark{b}  & $71.34$   & $3.07$  & $0.00$   & $0.00$  \\
\hline                                             %inserts single line
\end{tabular}
\tablefoot{
\tablefoottext{a}{The disk extent has been chosen according to a 2D grid scan for a total maximum likelihood over all 80 bins (see Sect.~\ref{sec:diff_comps}).}
\tablefoottext{b}{An extent of 0$^\circ$ is equivalent to a point source.}
\tablefoottext{c}{The Galactic Centre Source has been chosen to coincide with the position of Sgr A* (see text for details).}
}
\end{table}
% ----------------------------------------------------------------------

In total, we use $N_I=6$ components to model the celestial emission in this energy range, in addition to a two-component background model ($N_B=2$), described below. The celestial emission in this energy range is dominated by the bright 511~keV line emission from the Galaxy's centre, modelled by a narrow bulge (NB) and a broad bulge (BB), and the low surface-brightness disk. In the centre of the Milky Way, a point-like source, called Galactic Centre Source (GCS) was used to describe the morphology. The two strongest continuum sources in the sky, the Crab and Cygnus X-1, have been added to the sky models in order to improve the maximum likelihood fit (see Sect.~\ref{sec:conti_sources} for the significances of these point sources in the analysed energy range).

% ----------------------------------------------------------------------
\begin{figure}
  \centering
  \includegraphics[width=\linewidth]{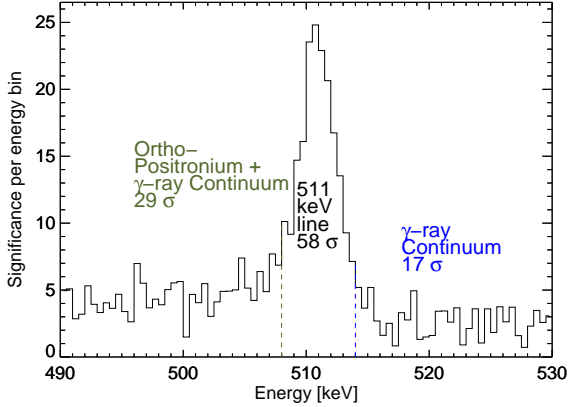}
   \caption{Detection of gamma-rays from the Galaxy in the spectral range of positron annihilation signatures. The significance of detecting a signal from the sky, summed over all spatial components as described in the text, is given for each 0.5~keV wide bin in units of $\sigma$. We can identify the intense line at 511~keV, together with an ortho-positronium continuum from positron annihilation, and an underlying gamma-ray continuum emission component.}
  \label{fig:SPI-spectrum_all}
\end{figure}
% ----------------------------------------------------------------------

In our model fitting we obtain model amplitudes in each of the 80 energy bins, for each of the sky model components, thus comprising an individual spectrum of celestial emission per component. Hence we are not biased towards any expected spectral shape from celestial emission. With the exception of exploring the effects of changing the disk parameters (which are most uncertain; see Sect.~\ref{sec:disk}), we do not alter the shape parameters of the image model components. We thus do not follow the optimisation of all sky components (i.e. NB, BB, CGS, and Disk) as studied in a single 6~keV wide energy bin, but use additional spectral information, and analyse cross correlation among components (see Appendix~\ref{sec:bulge_morph_effects}).

\subsection{Background modelling}
\label{sec:bg_model}

Our approach to instrumental background is not restricted to data in a specific energy bin, and exploits the physical nature of background. We study background lines with their characteristic shape, temporal changes, and relative intensities between detectors, and validate their behaviour on the basis of the associated physical processes. By combining data across a broader range of energy and time periods, we build a self-consistent description of spectral detector response and background characteristics, separately for continuum background and each line. At the same time, we ensure consistency of instrumental aspects (e.g. the resolution as it behaves with energy and time) as well as physical constraints (lines from the same isotopes behave identically). We thus separate long-term stable properties over the mission years from variations on shorter time scales caused by cosmic-ray intensity variations and detector degradation. Gamma-ray continuum and line backgrounds are treated separately, according to their different physical nature. 

We first determine what background lines are required to be taken into account by examining spectra accumulated over all detectors and the whole length of the data set. Then we apply spectral fitting to data accumulated over periods long enough for adequate detail in characteristic spectral signatures. Line shapes are thus determined for each detector over sufficiently long integration times to ensure that line centroids and widths are well defined beyond statistical uncertainties but only as long as the degradation becomes significant to change the line shape. The line shape $L(E)$ we use for the spectral fits is the convolution of a Gaussian function $G(E)$ with an exponential tail function towards lower energies $T(E)$, describing the degradation of each detector $j$, changing with time, superimposed on a power-law shaped continuum $C(E)$ (see Eqs.~(\ref{eq:conti})-(\ref{eq:fit_fun})). 
\begin{eqnarray}
C(E) & = & C_{0,j} \left( \frac{E}{511~\mathrm{keV}} \right)^{\alpha_{j}} \label{eq:conti} \\
G(E) & = & A_{0,ij} \exp \left(- \frac{(E-E_{0,ij})^2}{2\sigma_{ij}} \right) \label{eq:gaussian} \\
T(E) & = & \frac{1}{\tau_{ij}} \exp \left( - \frac{\tau_{ij}}{E} \right) \quad \forall E > 0 \label{eq:tail} \\
L(E) & = & (G \otimes T)(E) = \nonumber \\
& = & \sqrt{\frac{\pi}{2}} \frac{A_{0,ij} \sigma_{ij}}{\tau_{ij}} \exp \left( \frac{2 \tau_{ij} (E-E_{0,ij}) + \sigma_{ij}^2}{2 \tau_{ij}^2} \right) \nonumber \\
& & \erfc \left( \frac{\tau_{ij} (E-E_{0,ij}) + \sigma_{ij}^2}{\sqrt{2} \sigma_{ij} \tau_{ij}} \right)
\label{eq:fit_fun}
\end{eqnarray}
In Eqs.~(\ref{eq:conti})-(\ref{eq:fit_fun}), $C_{0,j}$ is the amplitude of the continuum at 511~keV for detector $j$, and $\alpha_{j}$ the respective power-law index. $A_{0,ij}$ is the amplitude of line $i$ in detector $j$, $E_{0,ij}$ is related to the peak value\footnote{The peak value of this convolved line shape is approximately $E_{peak} = E_0 - \tau$ for $\tau < 1~\mathrm{keV}$.} of the line shape, $\sigma_{ij}$ is the intrinsic width of a line in a particular detector, and $\tau_{ij}$ the degradation parameter in units of keV.

The time period used for accumulation, three days, is suitable (i.e. short enough) to investigate the gradual change in detector responses due to cosmic-ray bombardment. By fitting the integrated spectra per detector in the chosen time-intervals, we obtain a consistent database of background and detector response parameters ($C_{0,j}$, $\alpha_{j}$, $A_{0,ij}$, $E_{0,ij}$, $\sigma_{ij}$, $\tau_{ij}$) which provides the ingredients to re-build the instrumental background at each energy, time, and per each detector. Thereafter, we can model the changes between detector annealings\footnote{Due to cosmic ray bombardment, the lattice structure of the Ge detectors is damaged gradually, which causes deterioration of their spectral resolving properties. Therefore, twice a year, the detector array is heated up to $100~^\circ\mathrm{C}$ to repair the lattice structure of the Ge detectors. This is called annealing.} in the parameters describing the response of each detector as a linear function of time and energy.

The last step of background modelling is then the prediction of background for our dataset of interest, which we use to study celestial signals on top of the background, in each specific energy range of interest. The database parameters represent essentially the behaviour of the background dominated count rates in the instrument because the varying sky contribution is smeared out. The database parameters allow us to reconstruct a background pattern, i.e. the expected count rate of each detector relative to each others, for instrumental background lines and instrumental background continuum. Due to the nature of these processes, the pattern is different for these two background components, \emph{lines} and \emph{continuum}, for each analysed energy bin. The short-term pointing-to-pointing variations as traced by an instrumental rate (here: side shield assembly total rate of the SPI instrument, SSATOTRATE), are imprinted on top of the pre-defined patterns for a coherent description of the background due to cosmic rays.

Special care is, however, needed to obtain the proper absolute normalisation of this background model. The relative detector contributions to total continuum and line backgrounds may not be properly normalised, as each of these are derived from a data set stretching further in time. Therefore, we re-normalise the background model as re-built from the database again to the actual data, by fitting a time-dependent scaling parameter, $\theta_{i,t}$ in Eq.~(\ref{eq:model-fit}), per background component in addition to the proposed sky model scaling parameters.

In general, the detection of diffuse emission with a coded-mask telescope like SPI, i.e. emission on angular scales comparable to the FoV, relies on the correct comparison of flux in one pointing with that in another. Such emission will add only small variations from its pattern of relative detector ratios to the total signal. Furthermore, the average contribution from extended large-scale emission during the observations around a particular target on the sky will not vary much as the $5\times5$ dithering is performed. It will however likely change when the target direction of pointings is redirected by a significant fraction of the telescope FoV or more. The fixed relative detector patterns, derived for instrumental background \emph{lines} and instrumental background \emph{continuum}, are our key tools allowing the shadowgrams from the mask to be distinguished, as they change between pointings. Systematic mismatches of the celestial and background detector patterns may make it more difficult to find mask-encoded sky signals, and can thus reduce the sensitivity. We perform a re-normalisation of background detector patterns whenever the telescope is re-oriented to target a location in the sky that is more distant from the current than $\sim$ one FoV; re-scaling by fitting at three-day intervals for each of the two background components is adequate to recover proper normalisation.

\begin{figure}
  \centering
  \includegraphics[width=\linewidth]{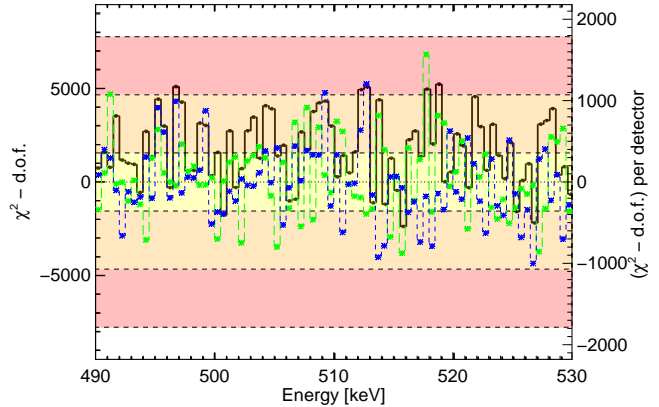}
   \caption{Background model performance measured by $\chi^2$-d.o.f. for each energy bin (i) for the entire SPI camera (left axis, black data points), and (ii) for two example detectors (right axis, detector 00 blue, detector 13 green). The ideal value of 0.0 (corresponding to a reduced $\chi^2$ of 1.0) is shown as a dotted line, together with the 1, 3, and 5$\sigma$ uncertainty intervals for a $\chi^2$-statistic with 1211021 d.o.f. The majority of points fall into the 3$\sigma$ band. No excess is evident either in the energy domain or for particular detectors.}
  \label{fig:SPI-BG}
\end{figure}

The adequacy of our background model has been assessed in Fig.~\ref{fig:SPI-BG}, showing $\chi^2-\mathrm{d.o.f.}$ for each energy bin. For the entire SPI-camera (black), the values scatter around a value of 1713 (corresponding to a reduced $\chi^2$ of $1.0014$ with 1211021 degrees of freedom (d.o.f.)) and fall well into a $3\sigma$ goodness-of-fit interval (orange area). No particular energy region is overemphasised in the maximum likelihood fits, nor are single detectors deviant. In total, our background model fitting determines 3772 parameters per energy bin.

% ----------------------------------------------------------------------
\begin{figure*}
	\centering
		\subfloat[Bulge. \label{fig:bulge}]{\includegraphics[width=0.50\textwidth]{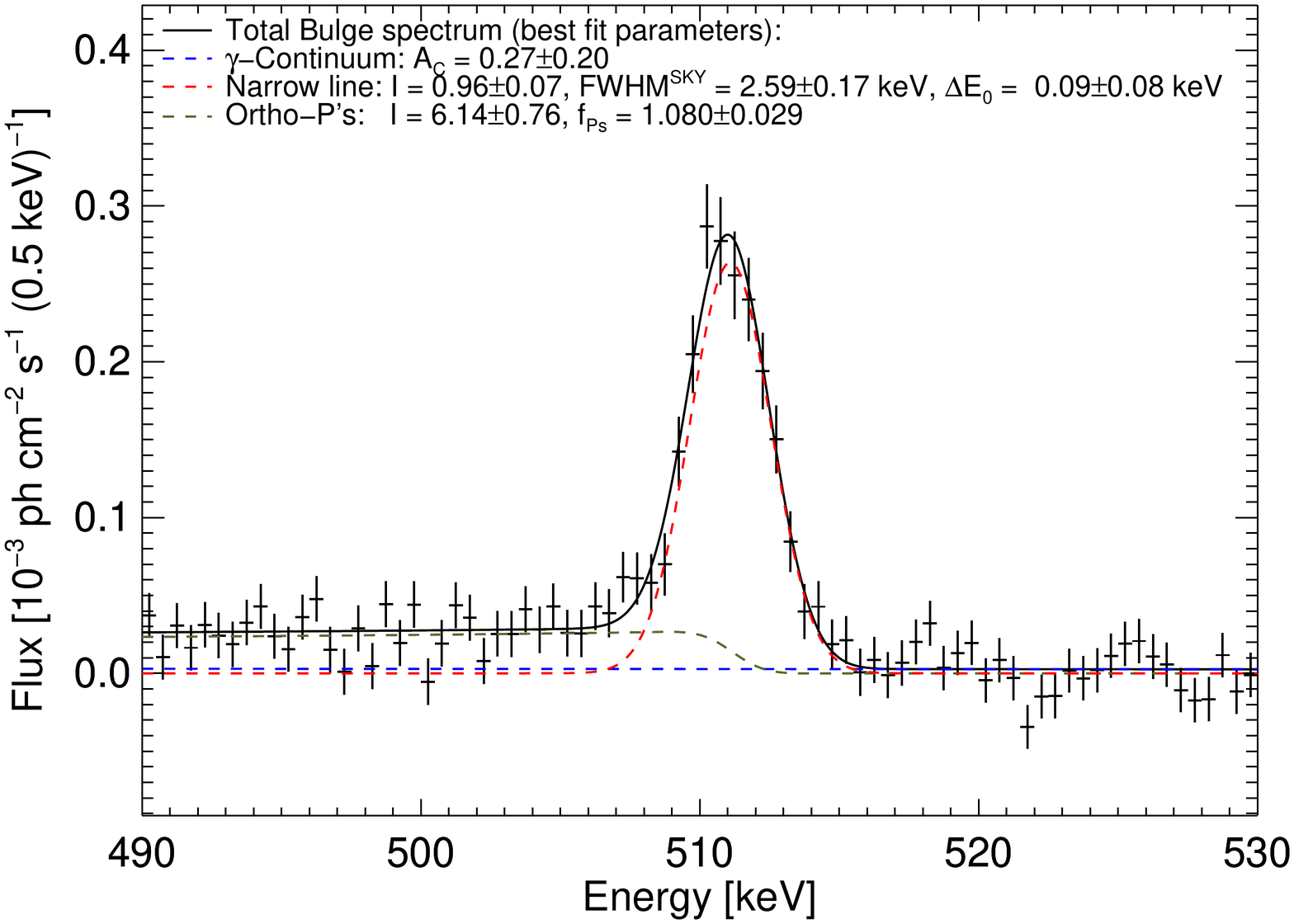}}
	  \subfloat[Disk. \label{fig:disk}]{\includegraphics[width=0.50\textwidth]{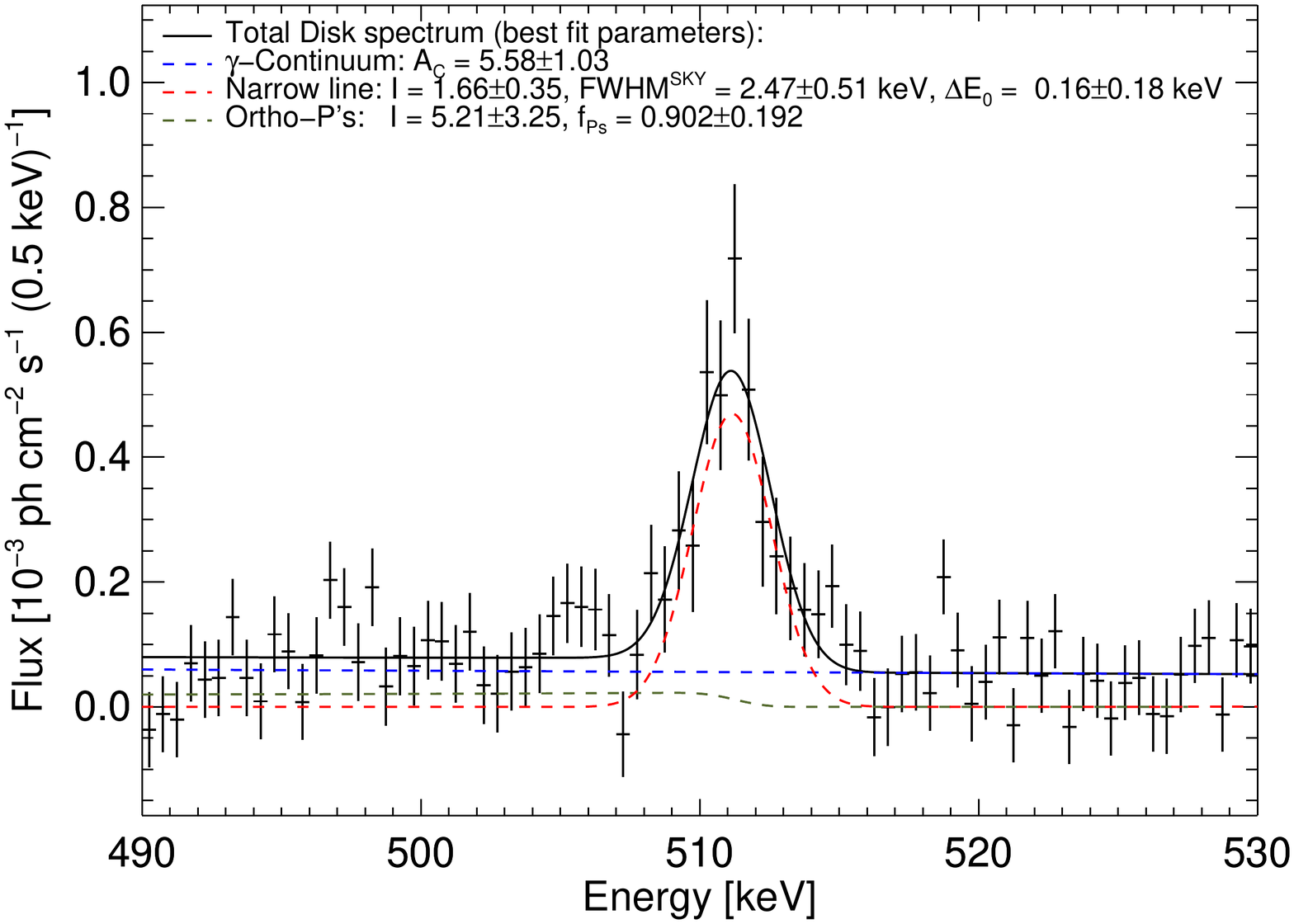}}
     \caption{Spectrum of annihilation gamma-rays from the bulge \emph{(a)} and the disk \emph{(b)}. The best fit spectrum is shown (continuous black line), as decomposed in a single 511~keV positron annihilation line (dashed red), the continuum from annihilation through ortho-positronium (dashed olive), and the diffuse gamma-ray continuum emission (dashed blue). Fitted and derived parameters are given in the legends. See text for details.}
     \label{fig:spi_spectra_bulge_disk}
\end{figure*}

% ----------------------------------------------------------------------
% ----------------------------------------------------------------------

%=======================================================
%
\section{Results and Interpretation}
\label{sec:results}

We determine spectra of sky emission from fits of intensity coefficients for each energy bin in a 40~keV wide range around the 511~keV line. 
Fig.~\ref{fig:SPI-spectrum_all} shows for each energy bin the detection significance of the celestial signal, based on our assumed description of the sky in terms of the six components in Tab.~\ref{tab:model_components} (see~Sect.~\ref{sec:sky_emission}). The significance is calculated using the likelihood-ratio between a background-only fit to the data and a fit including the background plus six celestial sources. The difference of d.o.f. in each energy bin is consequently 6. The dominant annihilation line at 511~keV is very clear with a total significance of 58$\sigma$, but characteristic ortho-positronium continuum on the low-energy side of the line is also detected with high significance, as is the underlying diffuse Galactic continuum emission~\citep{Bouchet2011_diffuseCR}. The continuum emission is predominantly from the disk and from the Crab and Cyg X-1.

\subsection{Different emission components}
\label{sec:diff_comps}

The  inner Galaxy is found to be the brightest region of the annihilation gamma-ray sky, confirming previous findings. It is detected with a significance of more than $56\sigma$~\citep[e.g.][]{Jean2006_511,Churazov2011_511}. We also detect a signal away from the inner Galaxy. Our disk-like emission component has a significance of $12\sigma$. The surface brightness of annihilation radiation for this disk component is rather low. The diffuse gamma-ray continuum emission from the Galaxy~\citep{Bouchet2011_diffuseCR} is the strongest signal in the disk, clearly detected even in this 40~keV band. The two strongest continuum sources in this energy band, the Crab and Cyg X-1 (point sources), are also detected, at $31\sigma$ and $11\sigma$, respectively; their spectral parameters are consistent with literature values (see Sect.\ref{sec:conti_sources}). In the centre of the Galaxy, an additional point-like source (or cusp, i.e. a point-like source that was recognised above the diffuse bulge emission to improve the overall fit to INTEGRAL observations in the 511~keV annihilation line) is needed to improve the fit. Fixing the positions and extents of the other components, we find a significance of $5\sigma$ for this component. 

% ----------------------------------------------------------------------
\begin{table}
\caption{Correlation coefficients for the six simultaneously fitted sky components.}              % title of Table
\label{tab:correlations}      % is used to refer this table in the text
\centering                                      % used for centering table
\begin{tabular}{c | r r r r r r}          % centered columns (6 columns)
\hline\hline                        % inserts double horizontal lines
%Name & Galactic Longitude Position [deg] & Galactic Latitude Position [deg] & Longitude extent (FWHM) [deg] & Latitude extent (FWHM) [deg] \\
             & NB        & BB        & Disk      & GCS       & Crab      & CX-1 \\
\hline                                   % inserts single horizontal line
    NB       & $1.000$  & $$        &  $$       & $$        & $$        & $$ \\      % inserting body of the table
    BB       & $-.836$  &  $1.000$  &  $$       & $$        & $$        & $$ \\
    Disk     & $.118$   & $-.365$   &  $1.000$  & $$        & $$        & $$ \\
    GCS      & $-.535$  &  $.224$   & $-.028$   &  $1.000$  & $$        & $$ \\
    Crab     & $-.018$  &  $.050$   & $-.102$   &  $.004$   &  $1.000$  & $$ \\
    CX-1  & $-.005$  &  $.003$   &  $.051$   &  $.001$   & $-.004$   &  $1.000$\\
\hline                                             %inserts single line
\end{tabular}
\tablefoot{Mean coefficients are given across all energy bins.}
\end{table}
% ----------------------------------------------------------------------

The signals from the different sky components cannot be determined independently, and we have calculated correlation coefficients for the values found for their intensities from the covariance matrix in the maximum likelihood fits. These are given in Tab.~\ref{tab:correlations}. Average values are given since the energy dependence is negligible, being less than 0.01\%.

We now discuss the results for each of the sky components.

% ----------------------------------------------------------------------
\subsubsection{The disk of the Galaxy}
\label{sec:disk}

% ----------------------------------------------------------------------
\begin{figure}
  \centering
  \includegraphics[width=\linewidth]{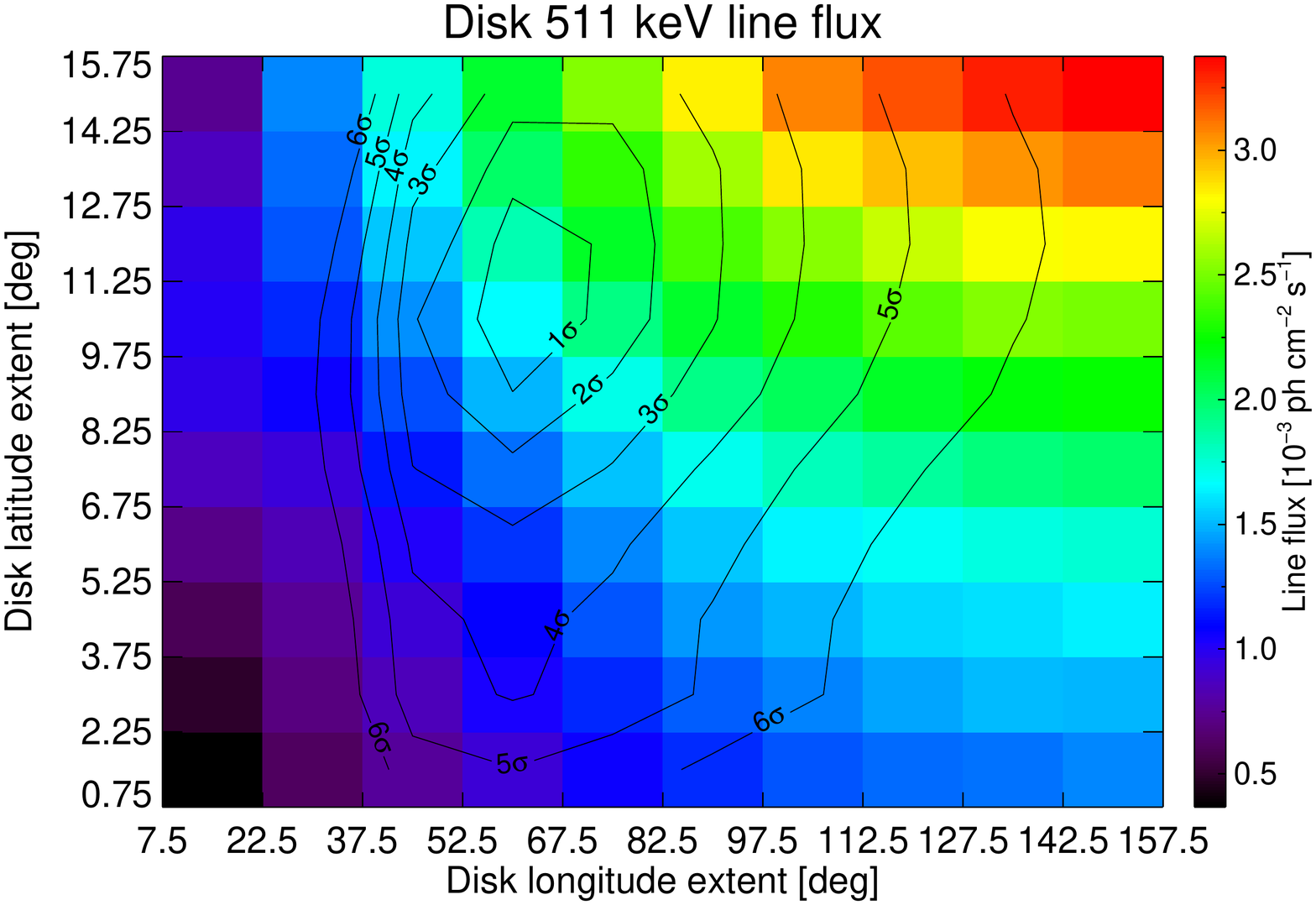}
  \includegraphics[width=\linewidth]{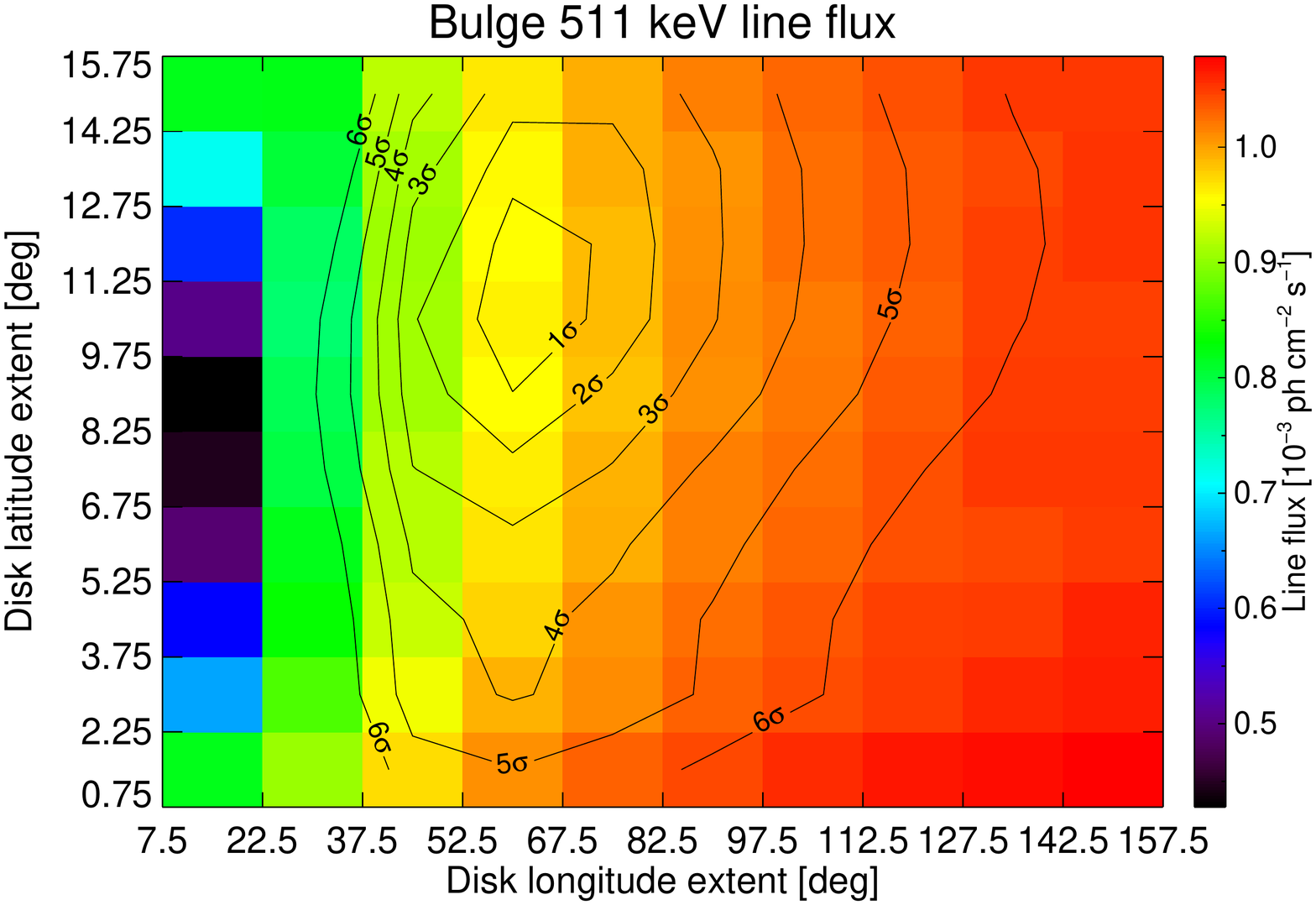}
 \caption{Dependence of the 511~keV line intensity in the disk (top) and the bulge (bottom) as a function of the choice of the disk  extent (1-$\sigma$ Gaussian width value). Line intensities are shown as shading, see scale on right-hand axis. Overlaid are the uncertainty contours for the disk size, as derived from the maximum likelihood fits in the grid-scan.  See Appendix~\ref{sec:cross_correlation_plots} for more details and other parameter impacts.}
  \label{fig:disk_ext_vs_disk_int}
\end{figure}
% ----------------------------------------------------------------------

The disk of the Galaxy is represented in our model by a two-dimensional Gaussian. We find a longitude extent of $60^{+10}_{-5}$~degrees, and a latitude extent of $10.5^{+2.5}_{-1.5}$~degrees (1-$\sigma$ values). In an independent analysis for the 511~keV line in a single 6~keV wide energy bin \citep{Skinner2014_511}, the disk extension parameters are found in the range of 30-90$^\circ$ in longitude and around 3$^\circ$ in latitude. Their study included the impact of different bulge parameters and different background methods. Assuming the background model described in Sect.~\ref{sec:bg_model} and the above mentioned bulge model, in our spectrally resolved analysis, we can reduce the uncertainty on the (model-dependent) disk extent.

Our  best disk extent is obtained by optimising for both the annihilation line and Galactic continuum components at the same time, for the energy region 490 to 530~keV. We find that the disk extent is also sensitive to how the central parts of the Galaxy are modelled, due to their overlap (see Appendix~\ref{sec:bulge_morph_effects} for further discussion of the bulge morphology dependence). Scanning solutions with different disk extents, given a fixed/optimal configuration of the bulge, and point sources, we find a maximum likelihood solution in each energy bin. In total, for a grid of 10$\times$10 different longitude and latitude widths, ranging from $\sigma_l=15^\circ,...,150^\circ$, and $\sigma_b=1.5^\circ,...,15^\circ$, respectively, models have been calculated for each of the 80 bins, and then fitted together with the other five sky model components and the two background model components. All model scaling parameters,  flux per energy bin and component, have been re-optimised for each point of this model grid. 

Fig.~\ref{fig:disk_ext_vs_disk_int} shows the dependence of the disk and the bulge 511~keV line intensity on the assumed disk-extent parameters. Contours indicate the uncertainty on the disk extent, as derived from the component-wise fit. As the disk becomes larger, its 511~keV line flux estimate increases because also very low surface brightness regions in the outer disk, and at high latitudes, contribute to the total flux (top). The individual relative uncertainties of each tile are almost constant at $\sim20\%$ for the disk, and $\sim5\%$ for the bulge. The bulge 511~keV line intensity is hardly dependent on the disk size (bottom). Over the disk longitude and latitude grid, the line flux changes by $\sim15\%$, whereas the $1\sigma$-uncertainty on the line-flux from this scan is essentially constant at $0.96\times10^{-3}~\mathrm{ph~cm^{-2}~s^{-1}}$ (see Sect.~\ref{sec:bulge} for further discussion). In the inner parts of the Galaxy ($|l|\lesssim45^\circ$), confusion between the bulge and the disk components causes the bulge to appear weaker compared to the disk.
The uncertainty of the intensity of the 511~keV line can be estimated from the tangents of equal flux as they intersect the $(2\Delta\log(L)=1)$-contours (where $L$ is the likelihood, see Appendix~\ref{sec:uncertainty_estimates} for details). A disk extent around 60$^\circ$ in longitude, and 10.5$^\circ$ in latitude, as suggested, does not agree with disk parameters obtained by other methods, focussing on the line only~\citep[e.g.][]{Bouchet2010_511}. The spectra per component retain their line shape and properties in a reasonably wide region around these best values, and systematic changes in total annihilation intensity remain small compared to statistical uncertainties.

Fig.~\ref{fig:disk} shows the disk spectrum for these optimum disk size parameters. 

We characterise these spectra in more detail, deriving the 511~keV line intensity ($I_L$), the width, characterised as kinematic broadening ($FWHM^{SKY}$), the centroid shift, interpreted as Doppler-shift from bulk motion ($\Delta E_0 = E_{peak} - E_{lab}$), the ortho-positronium intensity ($I_O$), and the positronium fraction ($f_{Ps}$). We represent the expected spectral components by a Gaussian 511 keV line, an ortho-positronium continuum~\citep{Ore1949_511}, and a power-law representing the diffuse Galactic gamma-ray continuum - each convolved with the SPI spectral response function and the parametrised kinematic broadening (see Sect.~\ref{sec:bg_model}). We use Monte Carlo sampling to determine the uncertainties of the fitted spectral characteristics, parametrised through the 511~keV line centroid, width, and amplitude, the ortho-positronium amplitude at the measured line centroid, and the continuum flux-density at 511~keV. As the power-law index for the diffuse Galactic continuum is poorly determined in our spectral band, and in any case has rather small impact on the annihilation component values ($\lesssim3\%$), we decided to fix its value a priori to $-1.7$~\citep{Kinzer1999_gamma,Kinzer2001_511,Strong2005_gammaconti,Jean2006_511,Churazov2011_511,Bouchet2011_diffuseCR}, consistent with our narrow-band fits. Likewise, the powerlaw index for the Crab and Cyg X-1 continua was set to $-2.23$ \citep{Jourdain2009_Crab}.  

We find a 511~keV line intensity for the disk of $(1.66\pm0.35)\times10^{-3}~\mathrm{ph~cm^{-2}~s^{-1}}$. Our Galactic gamma-ray continuum flux density of $(5.85\pm1.05)\times10^{-5}~\mathrm{ph~cm^{-2}~s^{-1}~keV^{-1}}$ at 511~keV corresponds to\footnote{Here we truncate the emission at 1\% of the maximal surface brighness; along the Galactic plane the intensity always is above that threshold and is therefore taken into account as $2\pi$; towards higher latitudes, 99\% of the emission are enclosed within $\sim70^\circ$. Hence, the disk emission encloses a solid angle of $3.11\pi~\mathrm{sr}$. Note, that this value is therefore model and threshold dependent.} $(5.99\pm1.07)\times10^{-6}~\mathrm{ph~cm^{-2}~s^{-1}~sr^{-1}~keV^{-1}}$ integrated across the full sky, consistent with results by~\citet{Strong2005_gammaconti} and \citet{Bouchet2011_diffuseCR} \footnote{\citet{Strong2005_gammaconti} and \citet{Bouchet2011_diffuseCR} focused on a broader energy range and on the central part of the Milky Way.}. 
The measured 511~keV line width in the disk is $(2.47\pm0.51)~\mathrm{keV}$ (FWHM). This is in concordance with the bulge value (see, however, the discussion in Sect.~\ref{sec:shapes}; see also Sect.~\ref{sec:bulge}). The 511~keV line shift, $(0.16\pm0.18)~\mathrm{keV}$, is consistent with zero. The ortho-positronium continuum has an intensity of $(5.21\pm3.25)\times10^{-3}~\mathrm{ph~cm^{-2}~s^{-1}}$.  

\begin{figure}
  \centering
  \includegraphics[width=\linewidth]{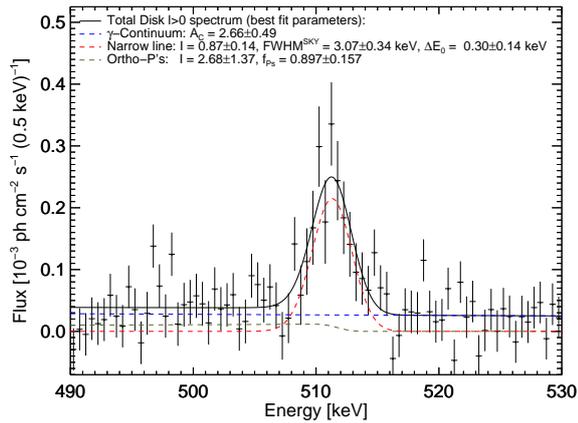} %{spec_disk_firstquad}
   \includegraphics[width=\linewidth]{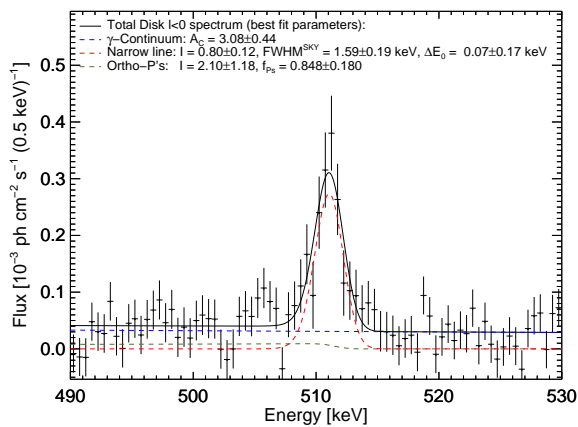} %{spec_disk_fourthquad}
 \caption{Spectrum of annihilation gamma rays from the eastern \emph{(top)}, and western \emph{(bottom)} hemisphere of the Galaxy's disk. The fitted parameters are given in the legends, colours are the same as in Fig.~\ref{fig:spi_spectra_bulge_disk}. No flux asymmetry is found.}
  \label{fig:spec_lr}
\end{figure}

Our statistics from eleven years of data allow us to derive spectral parameters separately for the eastern ($l>0^\circ$) and the western ($l<0^\circ$) hemisphere (see Figure~\ref{fig:spec_lr}). Here, the Gaussian-shaped disk component is masked on alternate sides ($l>0^\circ$, and $l<0^\circ$), which results in fitting now seven individual sky components. The 511~keV line intensities are $(0.87\pm0.14)\times10^{-3}~\mathrm{ph~cm^{-2}~s^{-1}}$ for the ($l>0^\circ$) region, and  $(0.80\pm0.12)\times10^{-3}~\mathrm{ph~cm^{-2}~s^{-1}}$ for the ($l<0^\circ$) region. Thus, we find no disk asymmetry in the line fluxes; in contrast to an earlier report~\citep{Weidenspointner2008a_511}, our east-west ratio is $1.09\pm0.24$. The asymmetry is reduced, if not completely removed, by shifting the narrow-bulge component away from the center by about $-1.25^\circ$ in longitude and $-0.25^\circ$ in latitude, as described above \citep[see][]{Skinner2014_511}. We find an east/west ratio for the ortho-positronium continuum of ($1.28\pm0.97$) and for the diffuse Galactic gamma-ray continuum of ($0.86\pm0.20$). All values are consistent with 1.0.

% ----------------------------------------------------------------------
% ----------------------------------------------------------------------

\subsubsection{The bulge region}
\label{sec:bulge}

We model the central part of the Milky Way by a combination of two 2D-Gaussians, NB and BB, from \citet{Skinner2014_511}. There, the NB is found to be centered at $(l,b)=(-1.25^\circ,-0.25^\circ)$ with Gaussian widths of $(\sigma_l,\sigma_b)=(2.5^\circ,2.5^\circ)$, the BB is found centered at $(l,b)=(0^\circ,0^\circ)$ and extends to $(\sigma_l,\sigma_b)=(8.7^\circ,8.7^\circ)$ (see Tab.~\ref{tab:model_components}). The correlation between the NB and BB components (see Tab.~\ref{tab:correlations}) with a value of $-0.836$ is not surprising, as two (or more) model components coincide/overlap spatially, when mapped onto the sky (see discussion in Sect.~\ref{sec:discussion} and Appendix~\ref{sec:cross_correlation_plots}). We define the \emph{bulge} component from the superposition of these two Gaussians, NB and BB. This definition represents an analytical description of an object's shape on the sky, independent and \emph{not} based on astronomical, model-biased definitions of \emph{the Galactic bulge}, e.g. as defined by a stellar population or by infrared emission. We prefer here this definition, as an alternative and complement to astrophysical model fitting, considering that such studies \citep{Weidenspointner2008a_511,Higdon2009_511} remained questionable.

In Fig.~\ref{fig:bulge}, the spectrum of the bright bulge shown. It has 511~keV line intensity of $(0.96\pm0.07)\times10^{-3}~\mathrm{ph~cm^{-2}~s^{-1}}$ and is detected with an overall significance of $56\sigma$. The bulge annihilation emission can be characterised by a 511~keV line of astrophysical width $(2.59\pm0.17)~\mathrm{keV}$, and a positronium fraction of (1.080$\pm$0.029), consistent with other recent studies~\citep{Jean2006_511,Churazov2011_511}. The line peak appears at $(511.09\pm0.08)~\mathrm{keV}$. The diffuse Galactic gamma-ray continuum is a minor contribution in the bulge; its intensity is $(0.27\pm0.20)\times10^{-5}~\mathrm{ph~cm^{-2}~s^{-1}~keV^{-1}}$.

% ----------------------------------------------------------------------
\begin{figure}
  \centering
   \includegraphics[width=\linewidth]{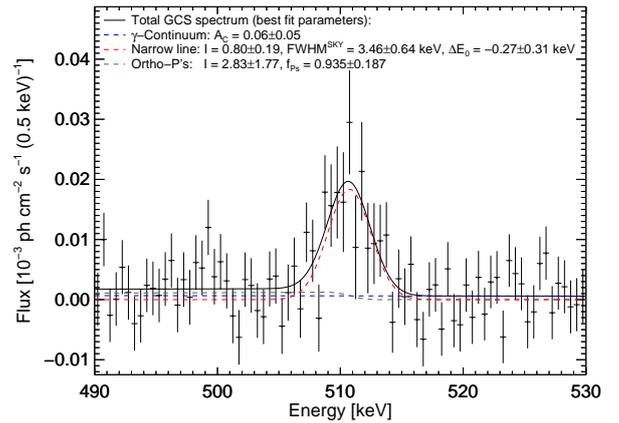}
 \caption{Spectrum of annihilation gamma rays from the point-like source (GCS) superimposed onto our extended bulge model in the Galaxy's centre. The fit and its components are indicated as above in Fig.~\ref{fig:spi_spectra_bulge_disk}.}
  \label{fig:SPI-spectra_central_component}
\end{figure}
% ----------------------------------------------------------------------

\subsubsection{The Galactic Centre region}
\label{sec:GCS_data}

The immediate vicinity in the direction of the centre of the Milky Way could be discriminated in \citet{Skinner2014_511} as a separate source or cusp. Given the spatial model that we adopt, the GCS is detected with a significance of 5$\sigma$ and we can provide a first spectrum from the annihilation emission of this source. The 511~keV line intensity is $(0.80\pm0.19)\times10^{-4}~\mathrm{ph~cm^{-2}~s^{-1}}$ (see Fig.~\ref{fig:SPI-spectra_central_component}). Its annihilation emission is characterised by a broadened line with a width of $(3.46\pm0.64)~\mathrm{keV}$ (FWHM above instrumental resolution), and a positronium fraction of (0.94$\pm$0.19). There is a slight hint of an underlying broad continuum with a flux density estimate of $(0.06\pm0.05)\times10^{-5}~\mathrm{ph~cm^{-2}~s^{-1}~keV^{-1}}$. There is no evident time variability down to scales of months.

The annihilation line is centred at $(510.73\pm0.31)~\mathrm{keV}$ if the spectrum is described by the above mentioned spectral model. Formally, there are indications of a red-shift: The ortho-positronium continuum is consistent with zero ($<2\sigma$), and assuming there is none leads to a slightly smaller value, $(510.59\pm0.35)~\mathrm{keV}$.

\subsubsection{Other sources}
\label{sec:conti_sources}

The Crab pulsar~\citep{Jourdain2009_Crab} and Cygnus X-1~\citep{Jourdain2012_CygX1} are the only known Galactic sources strong enough to influence our analysis and are included as constant point sources.

The Crab is detected in our energy band at 31$\sigma$ significance. Using a power-law with a fixed photon index\footnote{The energy range chosen is not enough to constrain the power-law index. The fitted value for the power-law index, given our data, is $(-1.41\pm1.52)$ for the Crab, and $(-2.9\pm4.5)$ for Cyg X-1. The resulting flux values change by less than 0.3\%.} of $-2.23$~\citep{Jourdain2009_Crab}, we find a flux density of $(2.20\pm0.07)\times10^{-5}~\mathrm{ph~cm^{-2}~s^{-1}~keV^{-1}}$ at 511~keV. Our overall flux in this energy band is consistent, though on the high side of, the analysis across the full energy range of SPI~\citep{Jourdain2009_Crab}. It is equivalent to about 40\% of the total diffuse Galactic gamma-ray continuum emission. 

Cygnus X-1 is detected at 11$\sigma$ significance. Its spectrum is described well by a single power-law spectrum with a (fixed) power-law index of $-2.23$, and a resulting flux density of $(0.65\pm0.06)\times10^{-5}~\mathrm{ph~cm^{-2}~s^{-1}~keV^{-1}}$. Cyg X-1 is time variable, with a hard, a soft, and possibly an intermediate state~\citep{McConnell2002_CygX1,Rodriguez2015_CygX1}. At 511~keV, the measured flux difference between the hard and the soft state is about $1.5\times10^{-4}~\mathrm{ph~cm^{-2}~s^{-1}~keV^{-1}}$~\citep{McConnell2002_CygX1}. Our measured average of the different possible spectral states of Cyg X-1 are in good agreement with recent measurements of~\citet{Rodriguez2015_CygX1}. As Cyg X-1 is only a weak source in the vicinity of 511~keV, and only used for improving the maximum likelihood fit, no time variability has been tested. 

The correlations between the continuum sources and the other sky model components are negligible, except for the correlation with the disk. In either case, and especially for Cyg X-1, the flux depends on the size of the disk emission model. If the disk is chosen to be (unrealistically) short, the flux density of Cyg X-1 captures (erroneously) a part of this disk emission.

We detect no ($<2\sigma$) annihilation signals in these additional sources. Upper limits are given for both sources in Tab.~\ref{tab:UL_cc}, assuming either a \emph{disk-like}, i.e. narrow 511~keV line ($1.6$~keV FWHM), or a \emph{GCS-like}, i.e. broad 511~keV line ($3.5$~keV FWHM), respectively.

% ----------------------------------------------------------------------
\begin{table}
\caption{Upper limits (2$\sigma$, in units $10^{-5}~\mathrm{ph~cm^{-2}~s^{-1}}$)  for 511~keV gamma-ray line emission originating from the Crab or Cyg X-1. Limits for a narrow ($1.6$~keV FWHM), and a broad line ($3.5$~keV FWHM) are given. }              % title of Table
\label{tab:UL_cc}      % is used to refer this table in the text
\centering                                      % used for centering table
\begin{tabular}{c | r r }          % centered columns (6 columns)
\hline\hline                        % inserts double horizontal lines
                & Narrow line   & Broad line         \\
\hline                                   % inserts single horizontal line
    Crab        & $6.62$        &  $7.44$   \\
    Cyg X-1     & $1.27$        &  $1.89$   \\
\hline                                             %inserts single line
\end{tabular}
\end{table}
% ----------------------------------------------------------------------

In addition to our six component model, we searched for possible point-like annihilation emission in the satellite galaxies of the Milky Way, because light dark matter particles may also be a candidate source of positrons \citep{Boehm2004_dm,Gunion2006_dm,Hooper2004_dm,Picciotto2005_dm,Pospelov2008_dm}, and dwarf galaxies are believed to be dark matter dominated~\citep[e.g.][]{Simon2007_dm,Strigari2008_dm}. We find $\geq 2\sigma$ excess emission in only six of 39 candidate sources (17 of 39 show a $1\sigma$ excess), which is insignificant considering the number of trials (Siegert et al. 2015, in prep.). 

We also searched for a 511 keV line from the Andromeda Galaxy (M31, modelled as a point-source), $(l/b)=(121.17/-21.57)^\circ$~\citep{Evans2010_chandra_cat}, finding an upper limit of $1\times10^{-4}~\mathrm{ph~cm^{-2}~s^{-1}}$ ($2\sigma$).

\section{Discussion}
\label{sec:discussion}

\subsection{Mutual spectral parameter compatibilities}
\label{sec:shapes}

% ----------------------------------------------------------------------
\begin{table*}
\caption{Spectral parameters for each sky component and respective $\chi^2$ fit values with d.o.f.. Continuum flux densities are given as the value at 511~keV in units of $10^{-5}~\mathrm{ph~cm^{-2}~s^{-1}~keV^{-1}}$, line and ortho-positronium fluxes are given in $10^{-4}~\mathrm{ph~cm^{-2}~s^{-1}}$, FWHM of the celestial emission line in keV, the centroid shift $\Delta E_0 = E_{peak} - E_{lab}$ in keV. The upper limits given for the line flux from the Crab and Cyg X-1 are $2\sigma$ values and the corresponding $\chi^2$ quoted are with no line. One sigma uncertainties are given in brackets.}              % title of Table
\label{tab:spec_deriv_params}      % is used to refer this table in the text
\centering                                      % used for centering table
%\begin{tabular}{c | r r r r r r r r r}          % centered columns (6 columns)
\begin{tabular}{l | >{\RaggedLeft}p{1.45cm} >{\RaggedLeft}p{1.25cm} >{\RaggedLeft}p{1.25cm} >{\RaggedLeft}p{1.45cm}  >{\RaggedLeft}p{1.45cm} >{\RaggedLeft}p{1.25cm} >{\RaggedLeft}p{1.65cm}}
\hline\hline                        % inserts double horizontal lines
%Name & Galactic Longitude Position [deg] & Galactic Latitude Position [deg] & Longitude extent (FWHM) [deg] & Latitude extent (FWHM) [deg] \\
                                       & Cont. flux dens. & Line Flux  & FWHM  &  $\Delta E_0$  & o-Ps Flux   & Pos. frac. & $\chi^2$/d.o.f.   \\
\hline                                   % inserts single horizontal line
    The Bulge                            & $0.27(20)$   & $9.6(7)$    & $2.59(17)$ & $ 0.09(8)$   & $61.4(7.6)$    & $1.08(3)$    & $66.47/74$ \\  
    Disk ($-180^\circ < l < 180^\circ$)  & $5.58(1.03)$ & $16.6(3.5)$ & $2.47(51)$ & $ 0.16(18)$  & $52.1(32.5)$   & $0.90(19)$   & $71.98/74$ \\
    Disk ($l>0^\circ$)                   & $2.66(49)$   & $8.7(1.4)$  & $3.07(34)$ & $ 0.30(14)$  & $26.8(13.7)$   & $0.90(16)$   & $83.79/74$ \\
    Disk ($l<0^\circ$)                   & $3.08(44)$   & $8.0(1.2)$  & $1.59(19)$ & $ 0.07(17)$  & $21.0(11.8)$   & $0.85(18)$   & $68.42/74$ \\
    GCS                                  & $0.06(5)$    & $0.8(2)$    & $3.46(64)$ & $-0.27(31)$  & $2.8(1.8)$     & $0.94(19)$   & $64.94/74$ \\
    Crab                                 & $2.20(7)$    & $<0.7$      & $-(-)$     & $-(-)$       & $-(-)$         & $-(-)$       & $66.97/78$ \\
    Cygnus X-1                           & $0.65(6)$    & $<0.2$      & $-(-)$     & $-(-)$       & $-(-)$         & $-(-)$       & $73.38/78$ \\
\hline
    Galaxy (total)                       & $8.79(85)$   & $27.4(3)$   & $2.61(23)$ & $0.15(9)$    & $116.3(29.3)$  & $0.99(7)$    & $75.00/74$ \\
\hline                                             %inserts single line
\end{tabular}
\end{table*}
% ----------------------------------------------------------------------

In Table~\ref{tab:spec_deriv_params}, we provide a summary of the fitted spectral parameters for all components included in our analysis, and the fitted parameters for a total, i.e. Galaxy-wide spectrum which serves as a conservative average of positron annihilation throughout the Milky Way. The line and continuum intensities of the single components add up to the total Galaxy-wide intensities, as expected. The widths of the single components formally range from 1.59 keV to 3.46 keV and are consistent with a mean value from the entire Galaxy except for the eastern and western hemisphere of the disk. The western hemisphere of the disk shows a smaller line width (FWHM of $(1.59\pm0.19)~\mathrm{keV}$) than the eastern hemisphere ($(3.07\pm0.34)~\mathrm{keV}$). We find a $2.8\sigma$ significance for the two halves to differ from each other. Each of the two halves differ by $1.4\sigma$ from the combined disk spectrum (see Appendix~\ref{sec:comp_spec} for further comparisons; see also Sect.~\ref{sec:annihilation_conditions}). Line widths above the instrumental resolution are in bulge and disk (total) $(2.59\pm0.17)$~keV and $(2.47\pm0.51)$~keV (FWHM), respectively, i.e. they are identical within uncertainties. Likewise, the line width of the bulge differs by $2.7\sigma$, compared to the line width of the western disk, and by $0.9\sigma$ to the eastern one. The GCS line width appears broader ($3.46\pm0.64$~keV) but is still consistent within $2\sigma$ uncertainties, compared to bulge and disk (total). The GCS is discussed in Sect.~\ref{sec:gcs_interpret}, separately. The Doppler-shifts, $\Delta E_0$, are essentially consistent with zero shift within $2\sigma$. The positron fractions are consistent with 1.0 throughout the galaxy (see below). The spectral fit quality is found adequate for all our model components.

\subsection{Annihilation conditions}
\label{sec:annihilation_conditions}
The positronium fraction, $f_{Ps}=2/(\frac{3}{2} + \frac{9}{4}\frac{I_L}{I_O})$, expresses how many positrons annihilate through an intermediate state of forming a positronium atom. It can be derived from the intensities of 511~keV line $I_L$ and ortho-positronium continuum $I_O$, and is a prime diagnostic of annihilation conditions. Formation of the positronium atom is only efficient below energies of $\sim$100~eV and is facilitated in a partially neutral medium through charge exchange reactions with atoms and molecules. In principle, the ionisation state, temperature, and composition (H, He, gas/dust) can be derived from the positronium fraction, comparing models and their predicted positronium fractions with values determined from spectral fits~\citep{Churazov2005_511,Jean2006_511,Guessoum2010_511,Churazov2011_511}. 

The measured positronium fractions in the bulge and in the disk as a whole are ($1.08\pm0.03$) and ($0.90\pm0.19$), respectively. Comparing the eastern and western hemisphere of the disk separately, we find values of ($0.90\pm0.16$) and ($0.85\pm0.18$), respectively. All values are consistent within uncertainties, and, moreover, with the theoretical physical limit of 1.0. Our choice of energy region (490 to 530~keV) may result in a bias towards high ortho-positronium flux. The total annihilation spectrum of the Galaxy shows a positronium fraction of ($0.99\pm0.07$). We conclude that the annihilation conditions based on only $f_{Ps}$ are the same throughout the entire disk and in the bulge within measurement uncertainties. 

A second order diagnostics of annihilation conditions is the shape, and particularly the width, of the 511~keV annihilation line. Here, kinematics of the positron population, and the gas to dust ratio of the ambient medium, are driving processes. The parameters which describe the annihilation conditions are listed in Tab.~\ref{tab:spec_deriv_params2} for all components.

% ----------------------------------------------------------------------
\begin{table}
\caption{Spectral parameters and converted physical properties for the main components. The values are quoted in units of $\mathrm{km~s^{-1}}$. One sigma uncertainties are given in brackets.}              % title of Table
\label{tab:spec_deriv_params2}      % is used to refer this table in the text
\centering                                      % used for centering table
%\begin{tabular}{c | r r r r r r r r r}          % centered columns (6 columns)
\begin{tabular}{l |  >{\RaggedLeft}p{1.65cm}  >{\RaggedLeft}p{1.65cm}  }
\hline\hline                        % inserts double horizontal lines
%Name & Galactic Longitude Position [deg] & Galactic Latitude Position [deg] & Longitude extent (FWHM) [deg] & Latitude extent (FWHM) [deg] \\
                                          & Velocity spread  &  Bulk motion   \\
\hline                                   % inserts single horizontal line
    The Bulge                             & $1500(100)$ & $53(47)$\\  
    Disk (total)                          & $1450(300)$ & $94(106)$ \\
    Disk $l>0^\circ$                      & $1800(200)$ & $176(82)$\\
    Disk $l<0^\circ$                      & $950(100)$  & $41(100)$\\
    GCS                                   & $2050(400)$ & $-159(182)$\\
 \hline
    Galaxy (total)                        & $1550(150)$ & $88(53)$  \\
 \hline                                             %inserts single line
\end{tabular}
\end{table}
% ----------------------------------------------------------------------

In the two disk hemispheres, we find a velocity spread of $(950\pm100)~\mathrm{km~s^{-1}}$ and $(1800\pm200)~\mathrm{km~s^{-1}}$, respectively, above the instrumental line width. This apparent difference in the line width in the two disk halves can be interpreted in several ways, depending on the relative contributions of kinematic and thermal broadening. The contribution from Galactic kinematics is probably small as the estimated velocity dispersion from interstellar gas is $\sim100~\mathrm{km~s^{-1}}$~\citep{Dame2001_galvel}, or up to $300~\mathrm{km~s^{-1}}$ if the positrons originate in the $\beta^+$-decay of $^{26}$Al in the disk~\citep{Kretschmer2013_26Al}. As we are performing line-of-sight integration over a whole hemisphere, we might be influenced by peculiar sampling of different annihilation regions, but then it would be surprising to observe the same flux.

\subsection{Sources throughout the Galaxy}
\label{sec:emission_sites}
In the bright bulge, the annihilation line is well represented by a single Gaussian-shaped 511~keV line, and the positronium continuum is clearly identified. In the disk of the Galaxy, the annihilation continuum from ortho-positronium is only marginally seen, the underlying Galactic-diffuse continuum is however clearly detected, while only marginal in the bulge region. We interpret this as a disk annihilation signature with its characteristic shape and rather homogeneous surface brightness, which is subdominant when viewing towards bulge due to the other components which have larger surface brightnesses. 

The bulge-to-disk (B/D) flux ratio is $0.58\pm0.13$; the (model-dependent\footnote{Based on effective distances to the bulge of $8.5~\mathrm{kpc}$, and to the disk of $10.0~\mathrm{kpc}$}) luminosity B/D ratio is $0.42\pm0.09$. As increasing exposure now reveals disk emission at low surface-brightness from an extended and thick disk, this may explain why our B/D ratio is lower than in previous analyses with less data, where B/D was reported >1 \citep{Knoedlseder2003_511,Knoedlseder2005_511,Weidenspointner2008a_511}. 
The bright bulge would preferably suggest origins of the positrons among old stellar populations, such as from SNe Ia, novae, LMXRBs, and microquasars~\citep[see discussion by][]{Prantzos2011_511}. Our best-fit latitude extent of the empirical disk model favours a rather large scale height ($\sim1~\mathrm{kpc}$). This suggests that positrons may be ejected from X-ray binaries and may annihilate further away from the sources, resulting in a low surface-brightness~\citep{Prantzos2008_511, Prantzos2011_511}. Accreting black-hole binaries may be more frequent in the bulge (3000 sources, \citet{Bandyopadhyay2009_511}) than in the disk, and could also reproduce the observed brightness distribution and disk extent. We have recently measured positron annihilation in the black hole binary V404 Cyg, which is in the Galactic disk , at $(l/b)=(73.12^\circ/-2.09^\circ)$, and $\sim 125~\mathrm{pc}$ above the Galactic plane, supporting the conjecture of microquasars as a source of Galactic positrons (Siegert et al. 2015, in review).

Positron production rates can be estimated from our measurements, assuming a steady state and effective source distances (see above). We obtain values of $2\times10^{43}~\mathrm{e^{+}~s^{-1}}$ and $3\times10^{43}~\mathrm{e^{+}~s^{-1}}$, for bulge and disk, respectively. These estimates are model-dependent, but help to discuss the order of magnitude for the positron production in the different components. 

Radioactivity from the $\beta^+$-decay of $^{26}$Al and $^{44}$Ti originating in massive stars contributes to positron production and 511~keV disk emission, as suggested from the $^{26}$Al morphology throughout the Galaxy derived with COMPTEL~\citep{Diehl1995_COMPTEL,Oberlack1996_26Al,Plueschke2001_26Al}, and recently with SPI~\citep{Bouchet2015_26Al}; see discussion by \citet{Martin2012_511}.

Massive stars ($^{26}$Al: $\dot{N}_{e^{+}} = 0.4\times10^{43}~\mathrm{e^{+}~s^{-1}}$), core-collapse supernovae ($^{44}$Ti: $\dot{N}_{e^{+}} = 0.3\times10^{43}~\mathrm{e^{+}~s^{-1}}$), LMXRBs ($\dot{N}_{e^{+}} \approx 1-2\times10^{43}~\mathrm{e^{+}~s^{-1}}$), and SNe Ia ($\dot{N}_{e^{+}} \approx 1-2\times10^{43}~\mathrm{e^{+}~s^{-1}}$) can add up to the total positron production rate in the Milky Way. But escape from sources and transport in the ISM surrounding the sources (escape fraction) is a key aspect for both SNe Ia and LMXRBs (in particular microquasars), which can result in major uncertainties in these absolute numbers~\citep{Alexis2014_511ISM}.

Detailed calculations of the positron escape fraction in the $^{56}$Ni$\rightarrow$$^{56}$Co$\rightarrow$$^{56}$Fe-decay chain, dominating the positron production in SNe Ia, predict $5\pm2$\% escape, based on the assumption of unmixed ejecta in W7 model (deflagration) calculations~\citep{Chan1993_511SNIa}. Observational support for this theoretical estimate has been found by subsequent studies~\citep{Milne1999_SNIa}, investigating the bolometric late-time light curves of a set of SNe Ia, with an average escape fraction value of $3.5\pm2$\% (see also \citet{Milne2001_SNIa} and the discussion by \citet{Kalemci2006_511}). From their $^{56}$Ni yield and Galactic type Ia supernova occurrence rate estimates, SNe Ia are expected to produce about $\dot{N}_{e^{+}} \approx (1.6\pm0.6)\times10^{43}~\mathrm{e^{+}~s^{-1}}$. Similar values have been found by others~\citep{Martin2012_511}, and thus would make SNe Ia through their $^{56}$Ni-decays one of the dominant positron producers in the Galaxy. Note that the positron escape fraction has not been measured directly, yet.

For microquasars, positron production has been proposed to arise from photon-photon interactions in their jets or closer to the compact gamma-ray source~\citep{Beloborodov1999_511,Guessoum2006_MQ511}, with a rough estimate of the average positron production rate of $\approx10^{41}~\mathrm{e^{+}~s^{-1}}$ per microquasar. Assuming the total measured annihilation rate of $(3.5-6)\times10^{43}~\mathrm{e^{+}~s^{-1}}$ to represent a steady state, it is necessary that several hundreds of microquasars are active at each time to meet the constraint. This is consistent with the expected/estimated number of black hole binaries in the Milky Way, between $10^3$ and a few $10^4$~\citep{Romani1992_LMXRB,Portegies_Zwart1997_LMXRB,Sadowski2008_LMXRB}, weighted with an average duty cycle (i.e. flaring (microquasar) vs. quiescent (XRB) state) between $10^{-3}-10^{-2}$. However, it is not known how many positrons annihilate directly in pair plasma ejecta or escape the source. A similar estimate was carried out by \citet{Weidenspointner2008a_511}, who suggested LMXRBs to be responsible for the asymmetric spatial distribution reported at that time. No such asymmetry is now observed.

The Fermi bubbles \citep{Su2012_jetsMW} might raise another type of candidate positron source: It has been suggested that the Fermi bubbles are manifestations of recent jet activity of Sgr A* \citep{Su2012_jetsMW,Yang2012_jetsFB}, and thus may be reminiscent of ordinary radio lobes in external galaxies \citep[see e.g.][]{Mingo2012_circinus}. There is evidence that such jet sources are powered by pair plasma \citep{Blandford1995_extragaljets,Konar2013_extragaljets,Hardcastle2015_extragaljets}. Jet dynamics would create a backflow in the lobes, which would drive the putative positrons back towards the Milky Way's disk \citep{Gaibler2011_extragaljets,Gaibler2012_extragaljets} where they would diffuse into the denser gas, annihilating as they are slowed down. For a total energy content of the Fermi bubbles of $10^{57}~\mathrm{erg}$ \citep{Yang2012_jetsFB}, and a 1\% share of this in positrons, the total positron reservoir in the Fermi bubbles of the order of $10^{59}$ positrons could sustain the Galactic annihilation emission for a few $10^{8}$ years.

\subsection{The Galactic Centre Source}
\label{sec:gcs_interpret}
The independent spectrum for the GCS is similar to that of the bulge and the disk spectrum. Considering the imaging resolution of SPI ($\sim$2.7$^\circ$), the source could be compact and related to the Galaxy's supermassive black hole Sgr A*, or encompass a region of the order of $350~\mathrm{pc}$ and thus include the entire central molecular zone (CMZ).

The supermassive black hole in the centre of the Milky Way with $(4.31\pm0.38)\times10^6~\mathrm{\Msol}$~\citep{Gillessen2009_SgraS2} and its associated accretion disk extending up to 100~AU~\citep[][and references therein]{Genzel2010_SgrA} have been discussed as positron sources \citep{Totani2006_accretion,Cheng2006_511}. Positrons can be produced in the vicinity of the supermassive black hole, from pair-production in the accretion disk, or in the hot corona above, or from the resulting jets~\citep{Beloborodov1999_accretion,Totani2006_accretion,Chernyshov2009_511}. If Sgr A* is the source of the positrons, we expect a gravitational red-shift of at least 0.4~keV, and a temperature of $\sim 10^3$~K~\citep{Shakura1973_accretion,Krolik1999_AGN}. These values are within the uncertainty limits of our measurement.

Alternatively, this Doppler broadening provides a measure of the turbulence of $(2000\pm400)~\mathrm{km~s^{-1}}$, which might reasonably be expected from non-equilibrium gas motion or past AGN activity. From its 511~keV luminosity of $(6.0\pm1.5)\times10^{41}~\mathrm{ph~s^{-1}}$, a steady state positron production rate of $(0.3$--$1.2)\times10^{42}~\mathrm{e^{+}~s^{-1}}$ is estimated, assuming positronium fractions between 0 and 1. Theoretical estimates are in the range of $0.16$--$3.7\times10^{42}~\mathrm{e^{+}~s^{-1}}$~\citep{Totani2006b_511}. 

\citet{Alexis2014_511ISM} discussed the possibility that nucleosynthesis positrons produced in the CMZ travel into the Galactic bulge and could be responsible for the emission in the extended bulge. It is, however, not implausible that these positrons annihilate locally in the CMZ. \citet{Alexis2014_511ISM} estimate the positron production from massive stars in the CMZ from $^{26}$Al as $0.3\times10^{42}~\mathrm{e^{+}~s^{-1}}$~\citep{Alexis2014_511ISM}. Adding the positrons produced by $^{44}$Ti, the observed $(1.0\pm0.5)\times10^{42}~\mathrm{e^{+}~s^{-1}}$ is within reach for a nucleosynthesis interpretation. Possibly the CMZ has now emerged as a separate positron annihilation site through our deeper exposure. 

The more pronounced peak of the 511~keV emission towards the centre of the Galaxy also may revive interpretations of a dark matter origin. Annihilating dark-matter particles have been proposed to create an annihilation photon emission profile proportional to the square of the dark matter density profile~\citep[e.g.][]{Burkert1995_dm,Navarro1996_dm,Merritt2006_dm} for the Milky Way~\citep[e.g.][]{Ascasibar2006_511dm}. The central cusp of such a distribution would be seen by SPI as a point-like source. As the nature of dark matter is entirely unknown, expected fluxes for the 511~keV line are based on rough estimates concerning the annihilation cross section, the dark matter particle mass and that the produced positrons do not propagate far. A comparison to theoretical estimates is therefore not possible. However, as satellite galaxies of the Milky Way are believed to be dark matter dominated, they should reveal a detectable 511~keV annihilation signal which is not consistently seen with SPI.

\section{Conclusions}
\label{sec:fazit}
Combining eleven years of INTEGRAL observations, we performed fine spectroscopy at 0.5~keV energy binning in the energy domain of the 511~keV line from positron annihilation. We obtain spectra for different regions of positron annihilation, which had been identified from earlier analysis, the bright bulge region and the disk of the Milky Way. The bulge region shows emission characteristic of a moderately warm and partly ionised interstellar gas, i.e. with 511~keV line and ortho-positronium continuum as clear signatures. There is an indication that the line width of bulge and part of the disk differ in detail ($2\sigma$). When the disk is separated into an eastern and a western hemisphere, the line widths from positive and negative longitudes also show a discrepancy at the $2\sigma$-level. In the disk, the ortho-positronium is hard to detect, because the disk surface-brightness of annihilation emission is fainter and the Galactic diffuse emission is relatively more intense. For the assumed bulge morphology and modelling approach, we find a disk extent of $60^{+10}_{-5}$~degrees in longitude, and $10.5^{+2.5}_{-1.5}$ in latitude (1-$\sigma$ values), corresponding to a scale height of $\sim$1~kpc. Given a small offset of the bright bulge, we do not find a flux asymmetry in the disk. The bulge-to-disk flux ratio is $0.58\pm0.13$, smaller than in earlier measurements. A combination of positrons from nucleosynthesis (radioactive $\beta^+$-decay), supernovae, and pair plasma ejection from accreting binary systems and possibly also a positron-rich backflow from past AGN-jet activity of Sgr A* appears consistent with our spectral and imaging results. In order to provide the bulk of positrons in the disk, one single source type is probably not efficient enough to provide the number of positrons observed. Other contributions cannot be excluded, as positron propagation in the interstellar medium and absolute positron ejection per source type remains uncertain. We have extracted separately the spectrum of a compact, possibly point-like, central region of the bulge and discussed its connection to possible sources, such as Sgr A*, the CMZ, and dark matter.

% ----------------------------------------------------------------------
%
\begin{acknowledgements}
  This research was supported by the German DFG cluster of excellence 'Origin and Structure of the Universe'. The INTEGRAL/SPI project has been completed under the responsibility and leadership of CNES; we are grateful to ASI, CEA, CNES, DLR, ESA, INTA, NASA and OSTC for support of this ESA space science mission. This work was also supported by funding from the Deutsche Forschungsgemeinschaft under DFG project number PR 569/10-1 in the context of the Priority Program 1573 Physics of the Interstellar Medium. The authors are grateful to Gerald K. Skinner for useful discussions.
 \end{acknowledgements}
% ----------------------------------------------------------------------
%
%\bibliographystyle{aa}
%\bibliography{alles}

\begin{thebibliography}{92}
\expandafter\ifx\csname natexlab\endcsname\relax\def\natexlab#1{#1}\fi

\bibitem[{{Albernhe} {et~al.}(1981){Albernhe}, {Le Borgne}, {Vedrenne},
  {Boclet}, {Durouchoux}, \& {da Costa}}]{Albernhe1981_511}
{Albernhe}, F., {Le Borgne}, J.~F., {Vedrenne}, G., {et~al.} 1981, \aap, 94,
  214

\bibitem[{{Alexis} {et~al.}(2014){Alexis}, {Jean}, {Martin}, \&
  {Ferri{\`e}re}}]{Alexis2014_511ISM}
{Alexis}, A., {Jean}, P., {Martin}, P., \& {Ferri{\`e}re}, K. 2014, \aap, 564,
  A108

\bibitem[{{Ascasibar} {et~al.}(2006){Ascasibar}, {Jean}, {B{\oe}hm}, \&
  {Kn{\"o}dlseder}}]{Ascasibar2006_511dm}
{Ascasibar}, Y., {Jean}, P., {B{\oe}hm}, C., \& {Kn{\"o}dlseder}, J. 2006,
  \mnras, 368, 1695

\bibitem[{{Bandyopadhyay} {et~al.}(2009){Bandyopadhyay}, {Silk}, {Taylor}, \&
  {Maccarone}}]{Bandyopadhyay2009_511}
{Bandyopadhyay}, R.~M., {Silk}, J., {Taylor}, J.~E., \& {Maccarone}, T.~J.
  2009, \mnras, 392, 1115

\bibitem[{{Beloborodov}(1999{\natexlab{a}})}]{Beloborodov1999_accretion}
{Beloborodov}, A.~M. 1999{\natexlab{a}}, in Astronomical Society of the Pacific
  Conference Series, Vol. 161, High Energy Processes in Accreting Black Holes,
  ed. J.~{Poutanen} \& R.~{Svensson}, 295

\bibitem[{{Beloborodov}(1999{\natexlab{b}})}]{Beloborodov1999_511}
{Beloborodov}, A.~M. 1999{\natexlab{b}}, \mnras, 305, 181

\bibitem[{{Blandford} \& {Levinson}(1995)}]{Blandford1995_extragaljets}
{Blandford}, R.~D. \& {Levinson}, A. 1995, \apj, 441, 79

\bibitem[{{Boehm} {et~al.}(2004){Boehm}, {Hooper}, {Silk}, {Casse}, \&
  {Paul}}]{Boehm2004_dm}
{Boehm}, C., {Hooper}, D., {Silk}, J., {Casse}, M., \& {Paul}, J. 2004,
  Physical Review Letters, 92, 101301

\bibitem[{{Bouchet} {et~al.}(2015){Bouchet}, {Jourdain}, \&
  {Roques}}]{Bouchet2015_26Al}
{Bouchet}, L., {Jourdain}, E., \& {Roques}, J.-P. 2015, \apj, 801, 142

\bibitem[{{Bouchet} {et~al.}(1991){Bouchet}, {Mandrou}, {Roques}, {Vedrenne},
  {Cordier}, {Goldwurm}, {Lebrun}, {Paul}, {Sunyaev}, {Churazov}, {Gilfanov},
  {Pavlinsky}, {Grebenev}, {Babalyan}, {Dekhanov}, \&
  {Khavenson}}]{Bouchet1991_mq511}
{Bouchet}, L., {Mandrou}, P., {Roques}, J.~P., {et~al.} 1991, \apjl, 383, L45

\bibitem[{{Bouchet} {et~al.}(2010){Bouchet}, {Roques}, \&
  {Jourdain}}]{Bouchet2010_511}
{Bouchet}, L., {Roques}, J.~P., \& {Jourdain}, E. 2010, \apj, 720, 1772

\bibitem[{{Bouchet} {et~al.}(2011){Bouchet}, {Strong}, {Porter}, {Moskalenko},
  {Jourdain}, \& {Roques}}]{Bouchet2011_diffuseCR}
{Bouchet}, L., {Strong}, A.~W., {Porter}, T.~A., {et~al.} 2011, \apj, 739, 29

\bibitem[{{Burkert}(1995)}]{Burkert1995_dm}
{Burkert}, A. 1995, \apjl, 447, L25

\bibitem[{{Chan} \& {Lingenfelter}(1993)}]{Chan1993_511SNIa}
{Chan}, K.-W. \& {Lingenfelter}, R.~E. 1993, \apj, 405, 614

\bibitem[{{Cheng} {et~al.}(2006){Cheng}, {Chernyshov}, \&
  {Dogiel}}]{Cheng2006_511}
{Cheng}, K.~S., {Chernyshov}, D.~O., \& {Dogiel}, V.~A. 2006, \apj, 645, 1138

\bibitem[{{Cheng} {et~al.}(1998){Cheng}, {Leventhal}, {Smith}, {Gehrels},
  {Tueller}, \& {Fishman}}]{Cheng1998_transient511}
{Cheng}, L.~X., {Leventhal}, M., {Smith}, D.~M., {et~al.} 1998, \apj, 503, 809

\bibitem[{{Chernyshov} {et~al.}(2009){Chernyshov}, {Cheng}, {Dogiel}, {Ko}, \&
  {Ip}}]{Chernyshov2009_511}
{Chernyshov}, D., {Cheng}, K.~S., {Dogiel}, V., {Ko}, C.~M., \& {Ip}, W.~H.
  2009, in The Extreme Sky: Sampling the Universe above 10 keV, 75

\bibitem[{{Churazov} {et~al.}(2011){Churazov}, {Sazonov}, {Tsygankov},
  {Sunyaev}, \& {Varshalovich}}]{Churazov2011_511}
{Churazov}, E., {Sazonov}, S., {Tsygankov}, S., {Sunyaev}, R., \&
  {Varshalovich}, D. 2011, \mnras, 411, 1727

\bibitem[{{Churazov} {et~al.}(2005){Churazov}, {Sunyaev}, {Sazonov},
  {Revnivtsev}, \& {Varshalovich}}]{Churazov2005_511}
{Churazov}, E., {Sunyaev}, R., {Sazonov}, S., {Revnivtsev}, M., \&
  {Varshalovich}, D. 2005, \mnras, 357, 1377

\bibitem[{{Dame} {et~al.}(2001){Dame}, {Hartmann}, \&
  {Thaddeus}}]{Dame2001_galvel}
{Dame}, T.~M., {Hartmann}, D., \& {Thaddeus}, P. 2001, \apj, 547, 792

\bibitem[{{Dermer} \& {Skibo}(1997)}]{Dermer1997_511}
{Dermer}, C.~D. \& {Skibo}, J.~G. 1997, \apjl, 487, L57+

\bibitem[{{Diehl}(1995)}]{Diehl1995_COMPTEL}
{Diehl}, R. 1995, Experimental Astronomy, 6, 103

\bibitem[{{Evans} {et~al.}(2010){Evans}, {Primini}, {Glotfelty}, {Anderson},
  {Bonaventura}, {Chen}, {Davis}, {Doe}, {Evans}, {Fabbiano}, {Galle}, {Gibbs},
  {Grier}, {Hain}, {Hall}, {Harbo}, {(Helen He}, {Houck}, {Karovska},
  {Kashyap}, {Lauer}, {McCollough}, {McDowell}, {Miller}, {Mitschang},
  {Morgan}, {Mossman}, {Nichols}, {Nowak}, {Plummer}, {Refsdal}, {Rots},
  {Siemiginowska}, {Sundheim}, {Tibbetts}, {Van Stone}, {Winkelman}, \&
  {Zografou}}]{Evans2010_chandra_cat}
{Evans}, I.~N., {Primini}, F.~A., {Glotfelty}, K.~J., {et~al.} 2010, \apjs,
  189, 37

\bibitem[{{Gaibler} {et~al.}(2011){Gaibler}, {Khochfar}, \&
  {Krause}}]{Gaibler2011_extragaljets}
{Gaibler}, V., {Khochfar}, S., \& {Krause}, M. 2011, \mnras, 411, 155

\bibitem[{{Gaibler} {et~al.}(2012){Gaibler}, {Khochfar}, {Krause}, \&
  {Silk}}]{Gaibler2012_extragaljets}
{Gaibler}, V., {Khochfar}, S., {Krause}, M., \& {Silk}, J. 2012, \mnras, 425,
  438

\bibitem[{{Genzel} {et~al.}(2010){Genzel}, {Eisenhauer}, \&
  {Gillessen}}]{Genzel2010_SgrA}
{Genzel}, R., {Eisenhauer}, F., \& {Gillessen}, S. 2010, Reviews of Modern
  Physics, 82, 3121

\bibitem[{{Gillessen} {et~al.}(2009){Gillessen}, {Eisenhauer}, {Fritz},
  {Bartko}, {Dodds-Eden}, {Pfuhl}, {Ott}, \& {Genzel}}]{Gillessen2009_SgraS2}
{Gillessen}, S., {Eisenhauer}, F., {Fritz}, T.~K., {et~al.} 2009, \apjl, 707,
  L114

\bibitem[{{Goldwurm} {et~al.}(1992){Goldwurm}, {Ballet}, {Cordier}, {Paul},
  {Bouchet}, {Roques}, {Barret}, {Mandrou}, {Sunyaev}, {Churazov}, {Gilfanov},
  {Dyachkov}, {Khavenson}, {Kovtunenko}, {Kremnev}, \&
  {Sukhanov}}]{Goldwurm1992_511}
{Goldwurm}, A., {Ballet}, J., {Cordier}, B., {et~al.} 1992, \apjl, 389, L79

\bibitem[{{Guessoum} {et~al.}(2005){Guessoum}, {Jean}, \&
  {Gillard}}]{Guessoum2005_511}
{Guessoum}, N., {Jean}, P., \& {Gillard}, W. 2005, \aap, 436, 171

\bibitem[{{Guessoum} {et~al.}(2010){Guessoum}, {Jean}, \&
  {Gillard}}]{Guessoum2010_511}
{Guessoum}, N., {Jean}, P., \& {Gillard}, W. 2010, \mnras, 402, 1171

\bibitem[{{Guessoum} {et~al.}(2006){Guessoum}, {Jean}, \&
  {Prantzos}}]{Guessoum2006_MQ511}
{Guessoum}, N., {Jean}, P., \& {Prantzos}, N. 2006, \aap, 457, 753

\bibitem[{{Guessoum} {et~al.}(1991){Guessoum}, {Ramaty}, \&
  {Lingenfelter}}]{Guessoum1991_511ISM}
{Guessoum}, N., {Ramaty}, R., \& {Lingenfelter}, R.~E. 1991, \apj, 378, 170

\bibitem[{{Gunion} {et~al.}(2006){Gunion}, {Hooper}, \&
  {McElrath}}]{Gunion2006_dm}
{Gunion}, J.~F., {Hooper}, D., \& {McElrath}, B. 2006, \prd, 73, 015011

\bibitem[{{Hardcastle}(2015)}]{Hardcastle2015_extragaljets}
{Hardcastle}, M. 2015, in Astrophysics and Space Science Library, Vol. 414,
  Astrophysics and Space Science Library, ed. I.~{Contopoulos}, D.~{Gabuzda},
  \& N.~{Kylafis}, 83

\bibitem[{{Harris} {et~al.}(1994){Harris}, {Share}, \&
  {Leising}}]{Harris1994_transient511}
{Harris}, M.~J., {Share}, G.~H., \& {Leising}, M.~D. 1994, \apj, 433, 87

\bibitem[{{Higdon} {et~al.}(2009){Higdon}, {Lingenfelter}, \&
  {Rothschild}}]{Higdon2009_511}
{Higdon}, J.~C., {Lingenfelter}, R.~E., \& {Rothschild}, R.~E. 2009, \apj, 698,
  350

\bibitem[{{Hooper} {et~al.}(2004){Hooper}, {Ferrer}, {Boehm}, {Silk}, {Paul},
  {Evans}, \& {Casse}}]{Hooper2004_dm}
{Hooper}, D., {Ferrer}, F., {Boehm}, C., {et~al.} 2004, Physical Review
  Letters, 93, 161302

\bibitem[{{Jean} {et~al.}(2009){Jean}, {Gillard}, {Marcowith}, \&
  {Ferri{\`e}re}}]{Jean2009_511ISM}
{Jean}, P., {Gillard}, W., {Marcowith}, A., \& {Ferri{\`e}re}, K. 2009, \aap,
  508, 1099

\bibitem[{{Jean} {et~al.}(2006){Jean}, {Kn{\"o}dlseder}, {Gillard}, {Guessoum},
  {Ferri{\`e}re}, {Marcowith}, {Lonjou}, \& {Roques}}]{Jean2006_511}
{Jean}, P., {Kn{\"o}dlseder}, J., {Gillard}, W., {et~al.} 2006, \aap, 445, 579

\bibitem[{{Jourdain} \& {Roques}(2009)}]{Jourdain2009_Crab}
{Jourdain}, E. \& {Roques}, J.~P. 2009, \apj, 704, 17

\bibitem[{{Jourdain} {et~al.}(2012){Jourdain}, {Roques}, \&
  {Malzac}}]{Jourdain2012_CygX1}
{Jourdain}, E., {Roques}, J.~P., \& {Malzac}, J. 2012, \apj, 744, 64

\bibitem[{{Jung} {et~al.}(1995){Jung}, {Kurfess}, {Johnson}, {Kinzer}, {Grove},
  {Strickman}, {Purcell}, {Grabelsky}, \& {Ulmer}}]{Jung1995_mq511}
{Jung}, G.~V., {Kurfess}, D.~J., {Johnson}, W.~N., {et~al.} 1995, \aap, 295,
  L23

\bibitem[{{Kalemci} {et~al.}(2006){Kalemci}, {Boggs}, {Milne}, \&
  {Reynolds}}]{Kalemci2006_511}
{Kalemci}, E., {Boggs}, S.~E., {Milne}, P.~A., \& {Reynolds}, S.~P. 2006,
  \apjl, 640, L55

\bibitem[{{Kinzer} {et~al.}(2001){Kinzer}, {Milne}, {Kurfess}, {Strickman},
  {Johnson}, \& {Purcell}}]{Kinzer2001_511}
{Kinzer}, R.~L., {Milne}, P.~A., {Kurfess}, J.~D., {et~al.} 2001, \apj, 559,
  282

\bibitem[{{Kinzer} {et~al.}(1999){Kinzer}, {Purcell}, \&
  {Kurfess}}]{Kinzer1999_gamma}
{Kinzer}, R.~L., {Purcell}, W.~R., \& {Kurfess}, J.~D. 1999, \apj, 515, 215

\bibitem[{{Kn{\"o}dlseder} {et~al.}(2005){Kn{\"o}dlseder}, {Jean}, {Lonjou},
  {Weidenspointner}, {Guessoum}, {Gillard}, {Skinner}, {von Ballmoos},
  {Vedrenne}, {Roques}, {Schanne}, {Teegarden}, {Sch{\"o}nfelder}, \&
  {Winkler}}]{Knoedlseder2005_511}
{Kn{\"o}dlseder}, J., {Jean}, P., {Lonjou}, V., {et~al.} 2005, \aap, 441, 513

\bibitem[{{Kn{\"o}dlseder} {et~al.}(2003){Kn{\"o}dlseder}, {Lonjou}, {Jean},
  {Allain}, {Mandrou}, {Roques}, {Skinner}, {Vedrenne}, {von Ballmoos},
  {Weidenspointner}, {Caraveo}, {Cordier}, {Sch{\"o}nfelder}, \&
  {Teegarden}}]{Knoedlseder2003_511}
{Kn{\"o}dlseder}, J., {Lonjou}, V., {Jean}, P., {et~al.} 2003, \aap, 411, L457

\bibitem[{{Konar} \& {Hardcastle}(2013)}]{Konar2013_extragaljets}
{Konar}, C. \& {Hardcastle}, M.~J. 2013, \mnras, 436, 1595

\bibitem[{{Kretschmer} {et~al.}(2013){Kretschmer}, {Diehl}, {Krause},
  {Burkert}, {Fierlinger}, {Gerhard}, {Greiner}, \&
  {Wang}}]{Kretschmer2013_26Al}
{Kretschmer}, K., {Diehl}, R., {Krause}, M., {et~al.} 2013, \aap, 559, A99

\bibitem[{{Krolik}(1999)}]{Krolik1999_AGN}
{Krolik}, J.~H. 1999, {Active galactic nuclei : from the central black hole to
  the galactic environment}

\bibitem[{{Leventhal} {et~al.}(1986){Leventhal}, {MacCallum}, {Huters}, \&
  {Stang}}]{Leventhal1986_511}
{Leventhal}, M., {MacCallum}, C.~J., {Huters}, A.~F., \& {Stang}, P.~D. 1986,
  \apj, 302, 459

\bibitem[{{Leventhal} {et~al.}(1978){Leventhal}, {MacCallum}, \&
  {Stang}}]{Leventhal1978_511}
{Leventhal}, M., {MacCallum}, C.~J., \& {Stang}, P.~D. 1978, \apjl, 225, L11

\bibitem[{{Lingenfelter} {et~al.}(2009){Lingenfelter}, {Higdon}, \&
  {Rothschild}}]{Lingenfelter2009_dm511}
{Lingenfelter}, R.~E., {Higdon}, J.~C., \& {Rothschild}, R.~E. 2009, Physical
  Review Letters, 103, 031301

\bibitem[{{Martin} {et~al.}(2012){Martin}, {Strong}, {Jean}, {Alexis}, \&
  {Diehl}}]{Martin2012_511}
{Martin}, P., {Strong}, A.~W., {Jean}, P., {Alexis}, A., \& {Diehl}, R. 2012,
  \aap, 543, A3

\bibitem[{{McConnell} {et~al.}(2002){McConnell}, {Zdziarski}, {Bennett},
  {Bloemen}, {Collmar}, {Hermsen}, {Kuiper}, {Paciesas}, {Phlips}, {Poutanen},
  {Ryan}, {Sch{\"o}nfelder}, {Steinle}, \& {Strong}}]{McConnell2002_CygX1}
{McConnell}, M.~L., {Zdziarski}, A.~A., {Bennett}, K., {et~al.} 2002, \apj,
  572, 984

\bibitem[{{Merritt} {et~al.}(2006){Merritt}, {Graham}, {Moore}, {Diemand}, \&
  {Terzi{\'c}}}]{Merritt2006_dm}
{Merritt}, D., {Graham}, A.~W., {Moore}, B., {Diemand}, J., \& {Terzi{\'c}}, B.
  2006, \aj, 132, 2685

\bibitem[{{Milne} {et~al.}(1999){Milne}, {The}, \& {Leising}}]{Milne1999_SNIa}
{Milne}, P.~A., {The}, L.-S., \& {Leising}, M.~D. 1999, \apjs, 124, 503

\bibitem[{{Milne} {et~al.}(2001){Milne}, {The}, \& {Leising}}]{Milne2001_SNIa}
{Milne}, P.~A., {The}, L.-S., \& {Leising}, M.~D. 2001, \apj, 559, 1019

\bibitem[{{Mingo} {et~al.}(2012){Mingo}, {Hardcastle}, {Croston}, {Evans},
  {Kharb}, {Kraft}, \& {Lenc}}]{Mingo2012_circinus}
{Mingo}, B., {Hardcastle}, M.~J., {Croston}, J.~H., {et~al.} 2012, \apj, 758,
  95

\bibitem[{{Navarro} {et~al.}(1996){Navarro}, {Frenk}, \&
  {White}}]{Navarro1996_dm}
{Navarro}, J.~F., {Frenk}, C.~S., \& {White}, S.~D.~M. 1996, \apj, 462, 563

\bibitem[{{Oberlack} {et~al.}(1996){Oberlack}, {Bennett}, {Bloemen}, {Diehl},
  {Dupraz}, {Hermsen}, {Knoedlseder}, {Morris}, {Schoenfelder}, {Strong}, \&
  {Winkler}}]{Oberlack1996_26Al}
{Oberlack}, U., {Bennett}, K., {Bloemen}, H., {et~al.} 1996, \aaps, 120, C311

\bibitem[{{Ore} \& {Powell}(1949)}]{Ore1949_511}
{Ore}, A. \& {Powell}, J.~L. 1949, Physical Review, 75, 1696

\bibitem[{{Picciotto} \& {Pospelov}(2005)}]{Picciotto2005_dm}
{Picciotto}, C. \& {Pospelov}, M. 2005, Physics Letters B, 605, 15

\bibitem[{{Pl{\"u}schke} {et~al.}(2001){Pl{\"u}schke}, {Diehl},
  {Sch{\"o}nfelder}, {Bloemen}, {Hermsen}, {Bennett}, {Winkler}, {McConnell},
  {Ryan}, {Oberlack}, \& {Kn{\"o}dlseder}}]{Plueschke2001_26Al}
{Pl{\"u}schke}, S., {Diehl}, R., {Sch{\"o}nfelder}, V., {et~al.} 2001, in ESA
  Special Publication, Vol. 459, Exploring the Gamma-Ray Universe, ed.
  A.~{Gimenez}, V.~{Reglero}, \& C.~{Winkler}, 55--58

\bibitem[{{Portegies Zwart} {et~al.}(1997){Portegies Zwart}, {Verbunt}, \&
  {Ergma}}]{Portegies_Zwart1997_LMXRB}
{Portegies Zwart}, S.~F., {Verbunt}, F., \& {Ergma}, E. 1997, \aap, 321, 207

\bibitem[{{Pospelov} {et~al.}(2008){Pospelov}, {Ritz}, \&
  {Voloshin}}]{Pospelov2008_dm}
{Pospelov}, M., {Ritz}, A., \& {Voloshin}, M. 2008, Physics Letters B, 662, 53

\bibitem[{{Prantzos}(2008)}]{Prantzos2008_511}
{Prantzos}, N. 2008, \nar, 52, 457

\bibitem[{{Prantzos} {et~al.}(2011){Prantzos}, {Boehm}, {Bykov}, {Diehl},
  {Ferri{\`e}re}, {Guessoum}, {Jean}, {Knoedlseder}, {Marcowith}, {Moskalenko},
  {Strong}, \& {Weidenspointner}}]{Prantzos2011_511}
{Prantzos}, N., {Boehm}, C., {Bykov}, A.~M., {et~al.} 2011, Reviews of Modern
  Physics, 83, 1001

\bibitem[{{Purcell} {et~al.}(1997){Purcell}, {Cheng}, {Dixon}, {Kinzer},
  {Kurfess}, {Leventhal}, {Saunders}, {Skibo}, {Smith}, \&
  {Tueller}}]{Purcell1997_511}
{Purcell}, W.~R., {Cheng}, L.-X., {Dixon}, D.~D., {et~al.} 1997, \apj, 491, 725

\bibitem[{{Purcell} {et~al.}(1993){Purcell}, {Grabelsky}, {Ulmer}, {Johnson},
  {Kinzer}, {Kurfess}, {Strickman}, \& {Jung}}]{Purcell1993_511}
{Purcell}, W.~R., {Grabelsky}, D.~A., {Ulmer}, M.~P., {et~al.} 1993, \apjl,
  413, L85

\bibitem[{{Rodriguez} {et~al.}(2015){Rodriguez}, {Grinberg}, {Laurent},
  {Cadolle Bel}, {Pottschmidt}, {Pooley}, {Bodaghee}, {Wilms}, \&
  {Gouiff{\`e}s}}]{Rodriguez2015_CygX1}
{Rodriguez}, J., {Grinberg}, V., {Laurent}, P., {et~al.} 2015, ArXiv e-prints

\bibitem[{{Romani}(1992)}]{Romani1992_LMXRB}
{Romani}, R.~W. 1992, \apj, 399, 621

\bibitem[{{Roques} {et~al.}(2003){Roques}, {Schanne}, {von Kienlin},
  {Kn{\"o}dlseder}, {Briet}, {Bouchet}, {Paul}, {Boggs}, {Caraveo},
  {Cass{\'e}}, {Cordier}, {Diehl}, {Durouchoux}, {Jean}, {Leleux}, {Lichti},
  {Mandrou}, {Matteson}, {Sanchez}, {Sch{\"o}nfelder}, {Skinner}, {Strong},
  {Teegarden}, {Vedrenne}, {von Ballmoos}, \& {Wunderer}}]{Roques2003_SPI}
{Roques}, J.~P., {Schanne}, S., {von Kienlin}, A., {et~al.} 2003, \aap, 411,
  L91

\bibitem[{{Sadowski} {et~al.}(2008){Sadowski}, {Zi{\'o}{\l}kowski},
  {Belczy{\'n}ski}, \& {Bulik}}]{Sadowski2008_LMXRB}
{Sadowski}, A., {Zi{\'o}{\l}kowski}, J., {Belczy{\'n}ski}, K., \& {Bulik}, T.
  2008, in American Institute of Physics Conference Series, Vol. 1010, A
  Population Explosion: The Nature \& Evolution of X-ray Binaries in Diverse
  Environments, ed. R.~M. {Bandyopadhyay}, S.~{Wachter}, D.~{Gelino}, \& C.~R.
  {Gelino}, 404--406

\bibitem[{{Shakura} \& {Sunyaev}(1973)}]{Shakura1973_accretion}
{Shakura}, N.~I. \& {Sunyaev}, R.~A. 1973, \aap, 24, 337

\bibitem[{{Share} {et~al.}(1988){Share}, {Kinzer}, {Kurfess}, {Messina},
  {Purcell}, {Chupp}, {Forrest}, \& {Reppin}}]{Share1988_511}
{Share}, G.~H., {Kinzer}, R.~L., {Kurfess}, J.~D., {et~al.} 1988, \apj, 326,
  717

\bibitem[{{Simon} \& {Geha}(2007)}]{Simon2007_dm}
{Simon}, J.~D. \& {Geha}, M. 2007, \apj, 670, 313

\bibitem[{{Skinner} {et~al.}(2014){Skinner}, {Diehl}, {Zhang}, {Bouchet}, \&
  {Jean}}]{Skinner2014_511}
{Skinner}, G., {Diehl}, R., {Zhang}, X., {Bouchet}, L., \& {Jean}, P. 2014, in
  Proceedings of the 10th INTEGRAL Workshop: ''A Synergistic View of the
  High-Energy Sky'' (INTEGRAL 2014). 15-19 September 2014. Annapolis, MD, USA.
  Published online at http://pos.sissa.it/cgi-bin/reader/conf.cgi?confid=228,
  id.054, 054

\bibitem[{{Skinner} {et~al.}(2010){Skinner}, {Jean}, {Kn{\"o}dlseder},
  {Martin}, {von Ballmoos}, \& {Weidenspointer}}]{Skinner2010_511}
{Skinner}, G., {Jean}, P., {Kn{\"o}dlseder}, J., {et~al.} 2010, in Eighth
  Integral Workshop. The Restless Gamma-ray Universe (INTEGRAL 2010), 20

\bibitem[{{Skinner} {et~al.}(2012){Skinner}, {Jean}, {Knoedlseder}, {von
  Ballmoos}, {Leising}, {Milne}, \& {Weidenspointner}}]{Skinner2012_511}
{Skinner}, G., {Jean}, P., {Knoedlseder}, J., {et~al.} 2012, in Proceedings of
  ''An INTEGRAL view of the high-energy sky (the first 10 years)'' - 9th
  INTEGRAL Workshop and celebration of the 10th anniversary of the launch
  (INTEGRAL 2012). 15-19 October 2012. Bibliotheque Nationale de France, Paris,
  France. Published online at
  http://pos.sissa.it/cgi-bin/reader/conf.cgi?confid=176, id.112, 112

\bibitem[{{Smith} {et~al.}(1996{\natexlab{a}}){Smith}, {Leventhal}, {Cavallo},
  {Gehrels}, {Tueller}, \& {Fishman}}]{Smith1996_transient511}
{Smith}, D.~M., {Leventhal}, M., {Cavallo}, R., {et~al.} 1996{\natexlab{a}},
  \apj, 471, 783

\bibitem[{{Smith} {et~al.}(1996{\natexlab{b}}){Smith}, {Leventhal}, {Cavallo},
  {Gehrels}, {Tueller}, \& {Fishman}}]{Smith1996_mq511crab}
{Smith}, D.~M., {Leventhal}, M., {Cavallo}, R., {et~al.} 1996{\natexlab{b}},
  \apj, 458, 576

\bibitem[{{Strigari} {et~al.}(2008){Strigari}, {Koushiappas}, {Bullock},
  {Kaplinghat}, {Simon}, {Geha}, \& {Willman}}]{Strigari2008_dm}
{Strigari}, L.~E., {Koushiappas}, S.~M., {Bullock}, J.~S., {et~al.} 2008, \apj,
  678, 614

\bibitem[{{Strong} {et~al.}(2005){Strong}, {Diehl}, {Halloin},
  {Sch{\"o}nfelder}, {Bouchet}, {Mandrou}, {Lebrun}, \&
  {Terrier}}]{Strong2005_gammaconti}
{Strong}, A.~W., {Diehl}, R., {Halloin}, H., {et~al.} 2005, \aap, 444, 495

\bibitem[{{Su} \& {Finkbeiner}(2012)}]{Su2012_jetsMW}
{Su}, M. \& {Finkbeiner}, D.~P. 2012, \apj, 753, 61

\bibitem[{{Sunyaev} {et~al.}(1991){Sunyaev}, {Churazov}, {Gilfanov},
  {Pavlinsky}, {Grebenev}, {Babalyan}, {Dekhanov}, {Khavenson}, {Bouchet},
  {Mandrou}, {Roques}, {Vedrenne}, {Cordier}, {Goldwurm}, {Lebrun}, \&
  {Paul}}]{Sunyaev1991_mq}
{Sunyaev}, R., {Churazov}, E., {Gilfanov}, M., {et~al.} 1991, \apjl, 383, L49

\bibitem[{{Totani}(2006)}]{Totani2006b_511}
{Totani}, T. 2006, \pasj, 58, 965

\bibitem[{{Totani} {et~al.}(2006){Totani}, {Oda}, {Sumi}, {Kosugi}, {Yasuda},
  \& {Doi}}]{Totani2006_accretion}
{Totani}, T., {Oda}, T., {Sumi}, T., {et~al.} 2006, in Astronomical Society of
  the Pacific Conference Series, Vol. 360, Astronomical Society of the Pacific
  Conference Series, ed. C.~M. {Gaskell}, I.~M. {McHardy}, B.~M. {Peterson}, \&
  S.~G. {Sergeev}, 55

\bibitem[{{Vedrenne} {et~al.}(2003){Vedrenne}, {Roques}, {Sch{\"o}nfelder},
  {Mandrou}, {Lichti}, {von Kienlin}, {Cordier}, {Schanne}, {Kn{\"o}dlseder},
  {Skinner}, {Jean}, {Sanchez}, {Caraveo}, {Teegarden}, {von Ballmoos},
  {Bouchet}, {Paul}, {Matteson}, {Boggs}, {Wunderer}, {Leleux},
  {Weidenspointner}, {Durouchoux}, {Diehl}, {Strong}, {Cass{\'e}}, {Clair}, \&
  {Andr{\'e}}}]{Vedrenne2003_SPI}
{Vedrenne}, G., {Roques}, J.-P., {Sch{\"o}nfelder}, V., {et~al.} 2003, \aap,
  411, L63

\bibitem[{{Weidenspointner} {et~al.}(2008){Weidenspointner}, {Skinner}, {Jean},
  {Kn{\"o}dlseder}, {von Ballmoos}, {Bignami}, {Diehl}, {Strong}, {Cordier},
  {Schanne}, \& {Winkler}}]{Weidenspointner2008a_511}
{Weidenspointner}, G., {Skinner}, G., {Jean}, P., {et~al.} 2008, \nat, 451, 159

\bibitem[{{Winkler} {et~al.}(2003){Winkler}, {Courvoisier}, {Di Cocco},
  {Gehrels}, {Gim{\'e}nez}, {Grebenev}, {Hermsen}, {Mas-Hesse}, {Lebrun},
  {Lund}, {Palumbo}, {Paul}, {Roques}, {Schnopper}, {Sch{\"o}nfelder},
  {Sunyaev}, {Teegarden}, {Ubertini}, {Vedrenne}, \&
  {Dean}}]{Winkler2003_INTEGRAL}
{Winkler}, C., {Courvoisier}, T.~J.-L., {Di Cocco}, G., {et~al.} 2003, \aap,
  411, L1

\bibitem[{{Yang} {et~al.}(2012){Yang}, {Ruszkowski}, {Ricker}, {Zweibel}, \&
  {Lee}}]{Yang2012_jetsFB}
{Yang}, H.-Y.~K., {Ruszkowski}, M., {Ricker}, P.~M., {Zweibel}, E., \& {Lee},
  D. 2012, \apj, 761, 185

\end{thebibliography}
%

%
\newpage
\begin{appendix} %First online appendix
\section{Cross correlations - impacts on spectral parameters}
\subsection{Estimating uncertainties}
\label{sec:uncertainty_estimates}
In our analysis, we model the sky distribution of gamma-ray emission through a predefined model set, composed of two bulge components, one disk component, and three point sources. This is clearly not an orthogonal model set, and thus dependencies occur between component results. We must estimate the impact of our choices of sky emission morphology on the spectral properties, accounting for cross correlations that will depend on the magnitude of the overlap / degeneracy among the sky model components.

\begin{figure}[!ht]
  \centering
  \includegraphics[width=\linewidth]{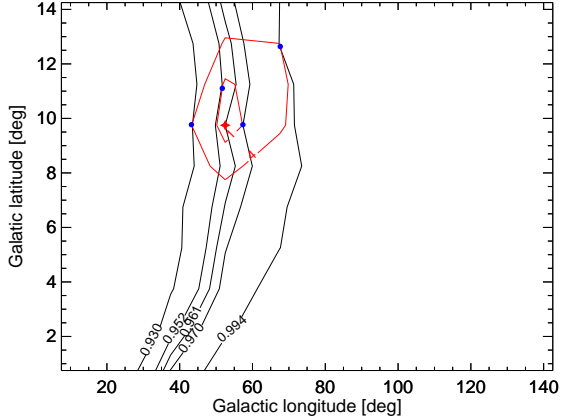} 
 \caption{511~keV line flux in the bulge, dependent on the disk size (analogous to Fig.~\ref{fig:disk_ext_vs_disk_int}). Shown are the lines of equal flux touching the $(2\Delta\log(L)=1)$-, and $(2\Delta\log(L)=4)$-contours, respectively, marked by the blue dots. These tangents correspond to the 1 and $2\sigma$ uncertainties of the line flux in the bulge with respect to the disk size (longitude and latitude extent). The star symbol is marking the point with the largest likelihood at $0.961\cdot10^{-3}\mathrm{~ph~cm^{-2}~s^{-1}}$. The resulting 1 and $2\sigma$-uncertainties are $(0.961\pm0.009)\cdot10^{-3}\mathrm{~ph~cm^{-2}~s^{-1}}$, and $(0.961^{+0.033}_{-0.031})\cdot10^{-3}\mathrm{~ph~cm^{-2}~s^{-1}}$, respectively.}
  \label{fig:extract_error}
\end{figure}

In Figs.~\ref{fig:spi_spectra_bulge_disk}, \ref{fig:spec_lr}, and \ref{fig:SPI-spectra_central_component}, showing the different spectra, the uncertainties on fitted parameters are based on the derived spectral data points and their respective uncertainties, given an optimised emission model. The uncertainties on the fitted parameters depend on parameter assumptions, e.g. also on the assumed bulge and disk size. Utilizing the plots of Sect.~\ref{sec:cross_correlation_plots}, the uncertainties for each fitted \emph{spectral} parameter dependent on the disk size can be estimated by the tangents of equal flux, line width, etc., touching the $(2\Delta\log(L)=1)$-contours, where $L$ is the likelihood. In Fig.~\ref{fig:extract_error}, this procedure is illustrated for the 511~keV line flux of the bulge. Other parameter uncertainties can be extracted analogously.

\subsection{Merging the bulge spectra}
The spectrum of \emph{the bulge} in Fig.~\ref{fig:bulge} is a superposition of the derived spectra of the NB and BB spectra. Given the significant correlation between NB and BB, the propagation of uncertainties takes the covariance between the components into account:
\begin{equation}
\sigma_{Bulge,k} = \sqrt{\theta_{0,k}^{2} \sigma_{0,k}^{2} + \theta_{1,k}^{2} \sigma_{1,k}^{2} + 2 \theta_{0,k} \theta_{1,k} \sigma_{01,k}}
\label{eq:gauss_prop_cova}
\end{equation}
In Eq.~(\ref{eq:gauss_prop_cova}), $\sigma_{Bulge,k}$ is the propagated uncertainty in energy bin $k$, as shown in Fig.~\ref{fig:bulge}, $\sigma_{i,k}$ are the uncertainties for components $i=0$ (NB), and $i=1$ (BB), in energy bin $k$ from the maximum-likelihood fit, Eq.~(\ref{eq:model-fit}). $\sigma_{01,k}$ is the covariance between the NB and the BB for each energy bin. Neglecting the covariance term would overestimate the statistical uncertainties by about 250\%.

\subsection{Disk size effects}
\label{sec:cross_correlation_plots}
\begin{figure*}
  \centering
  \includegraphics[width=0.33\linewidth]{Bulge_line_int.pdf}
  \includegraphics[width=0.33\linewidth]{Disk_line_int.pdf}
  \includegraphics[width=0.33\linewidth]{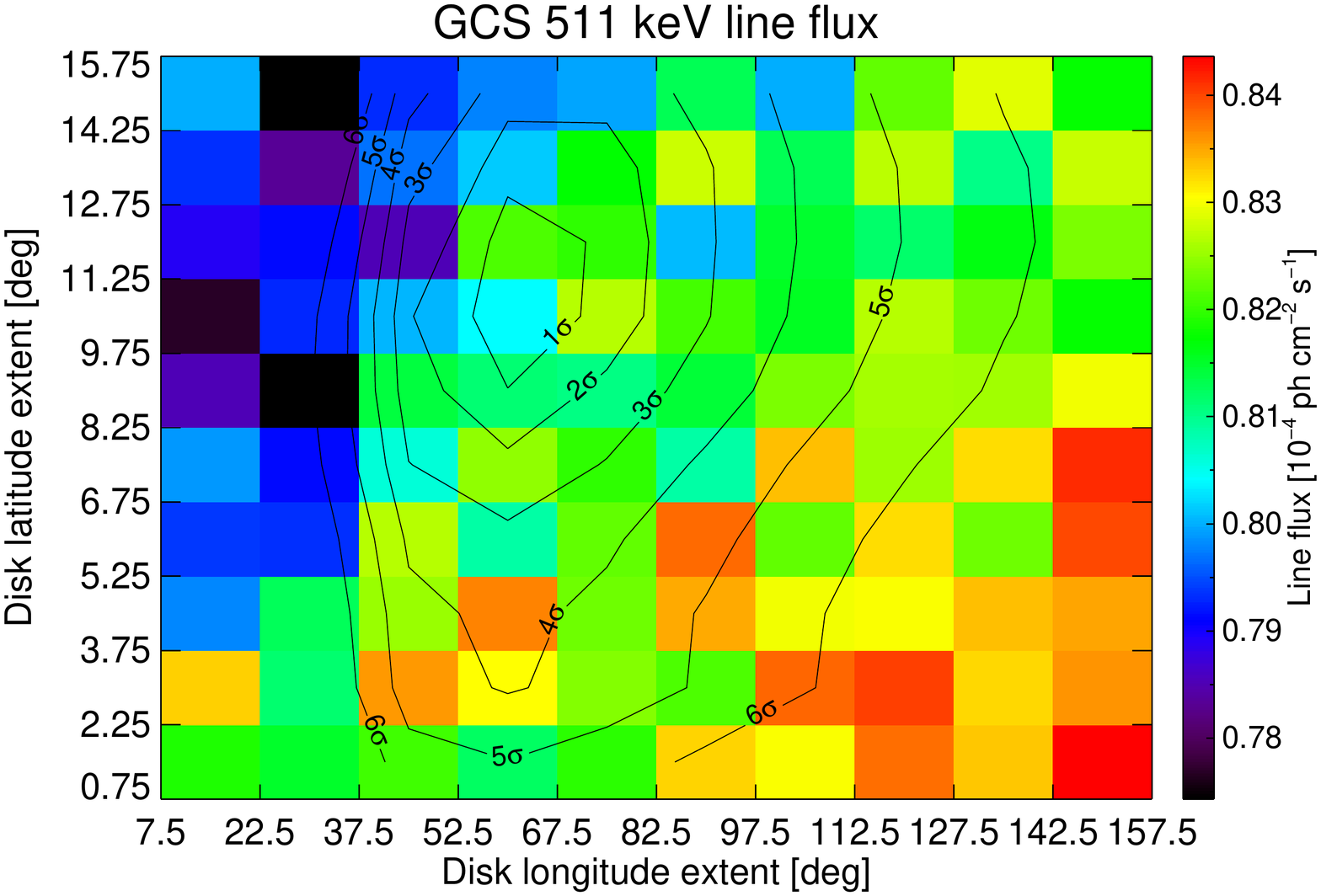}
  \\
  \includegraphics[width=0.33\linewidth]{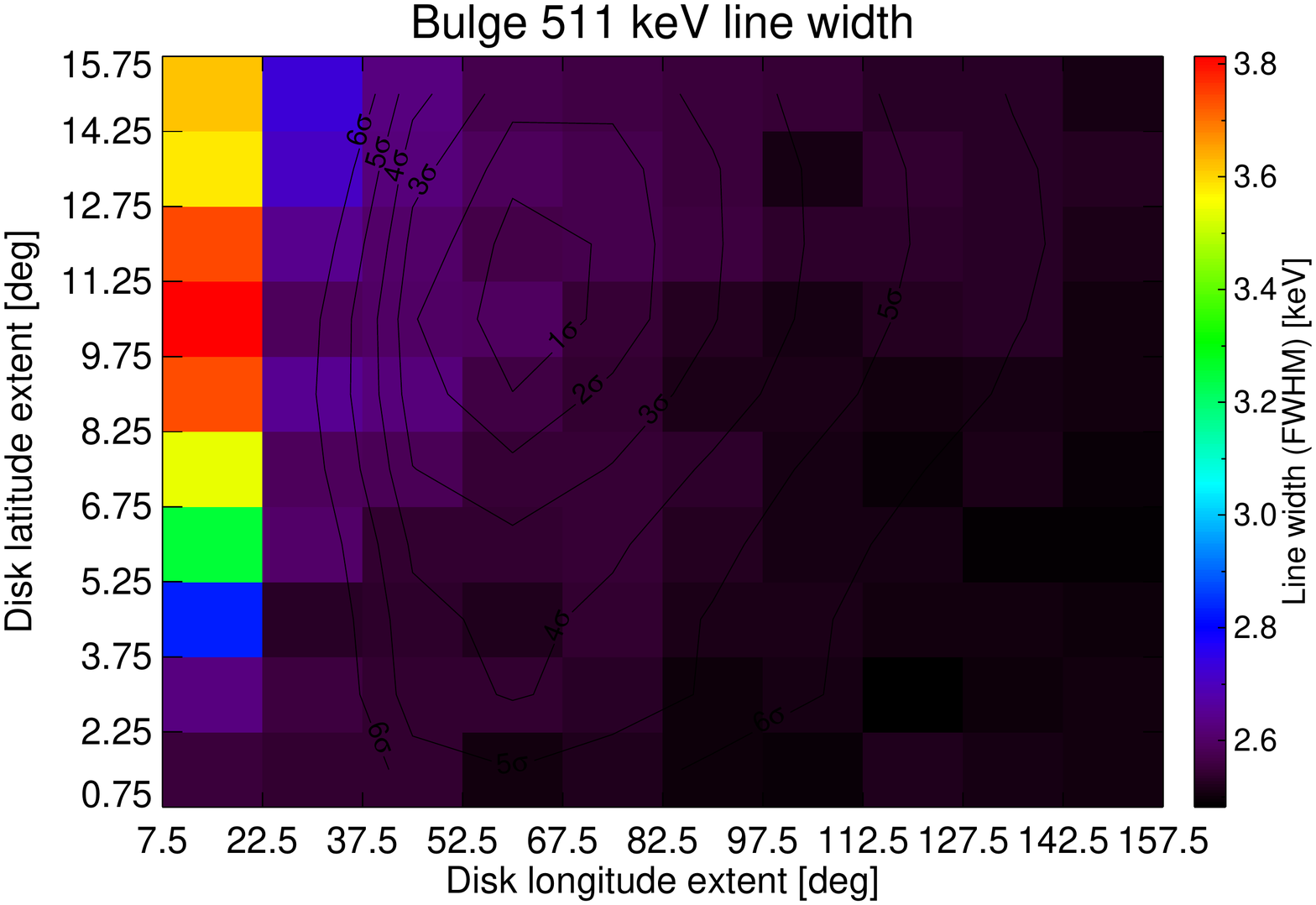}
  \includegraphics[width=0.33\linewidth]{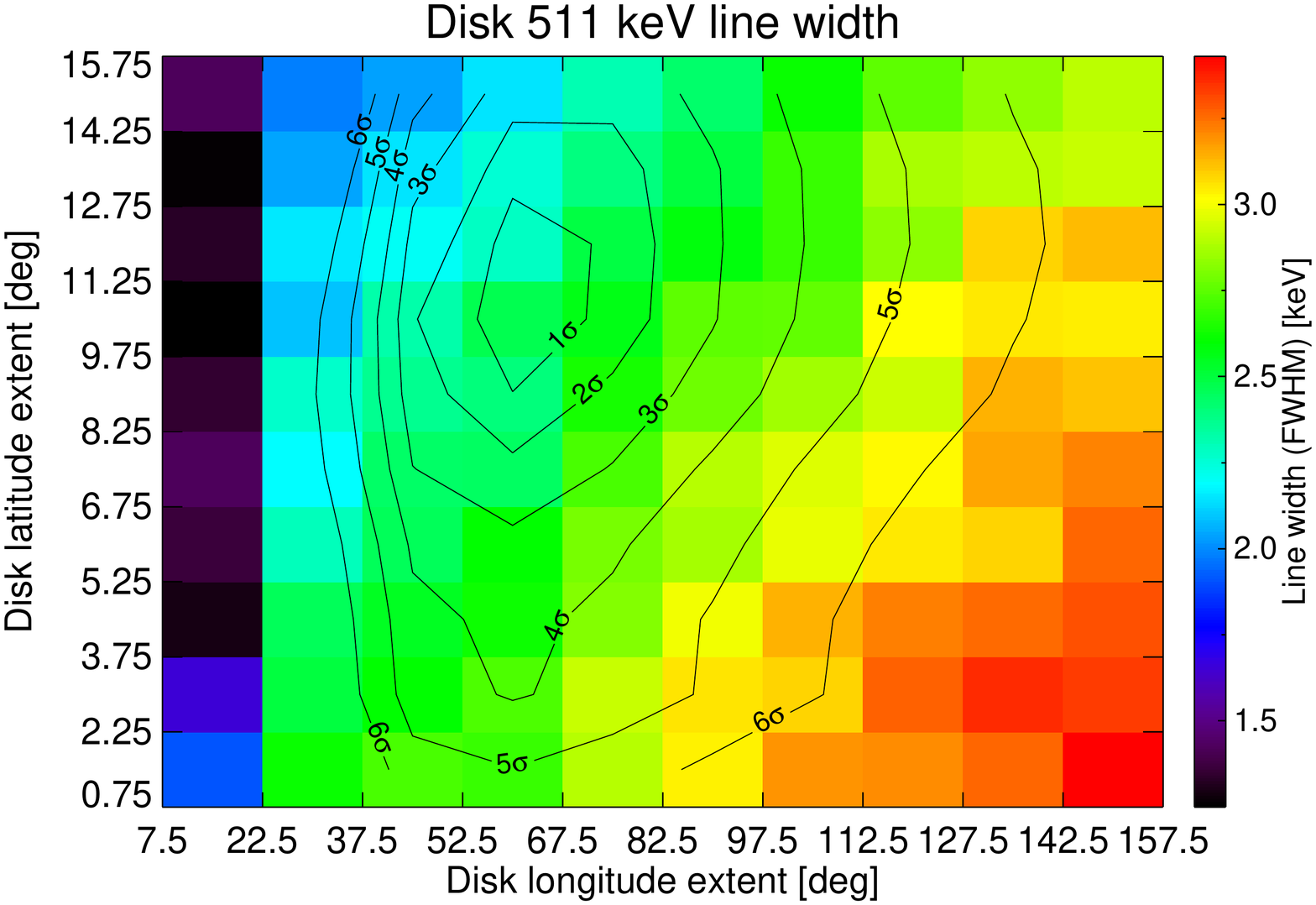}
  \includegraphics[width=0.33\linewidth]{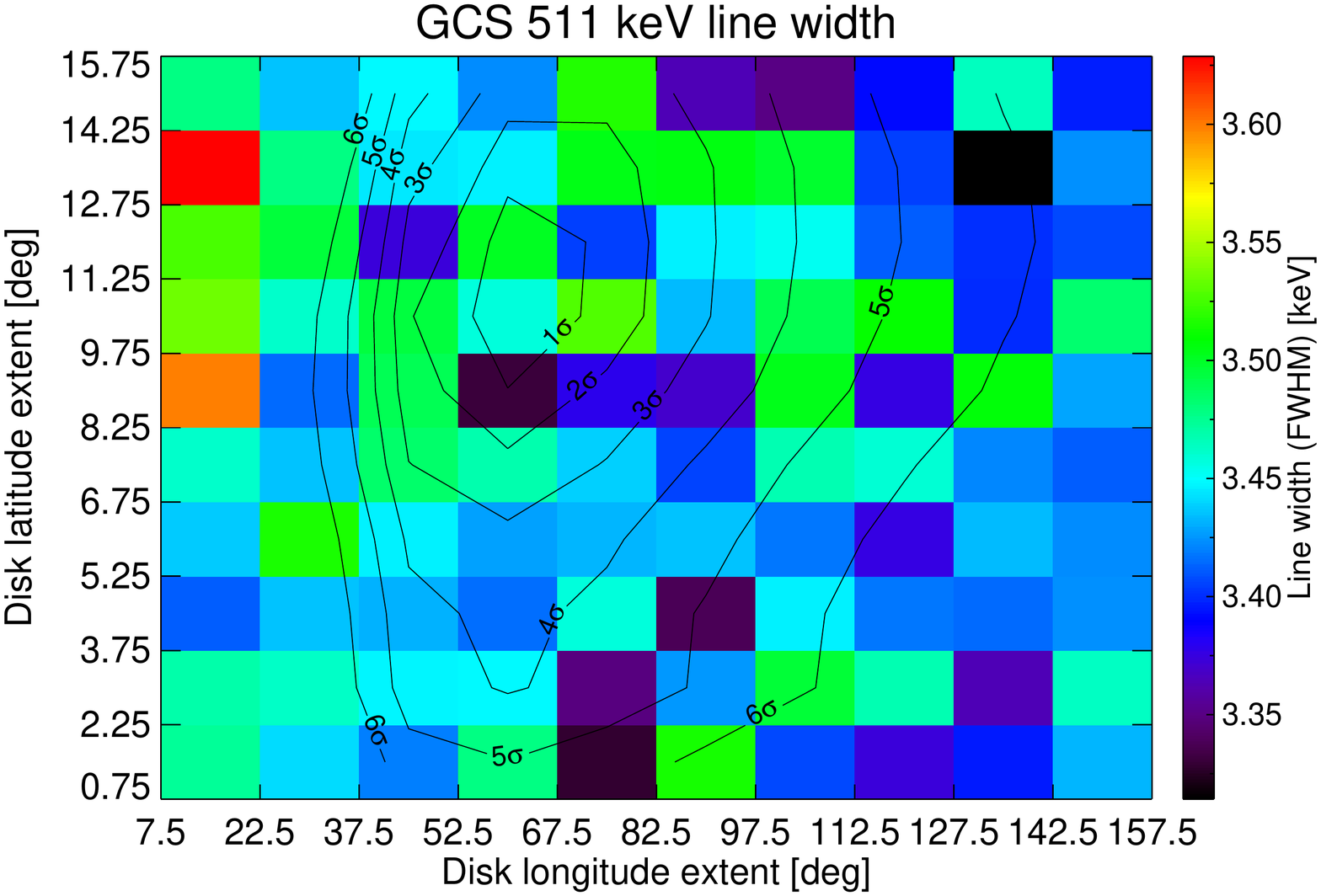}
  \\
  \includegraphics[width=0.33\linewidth]{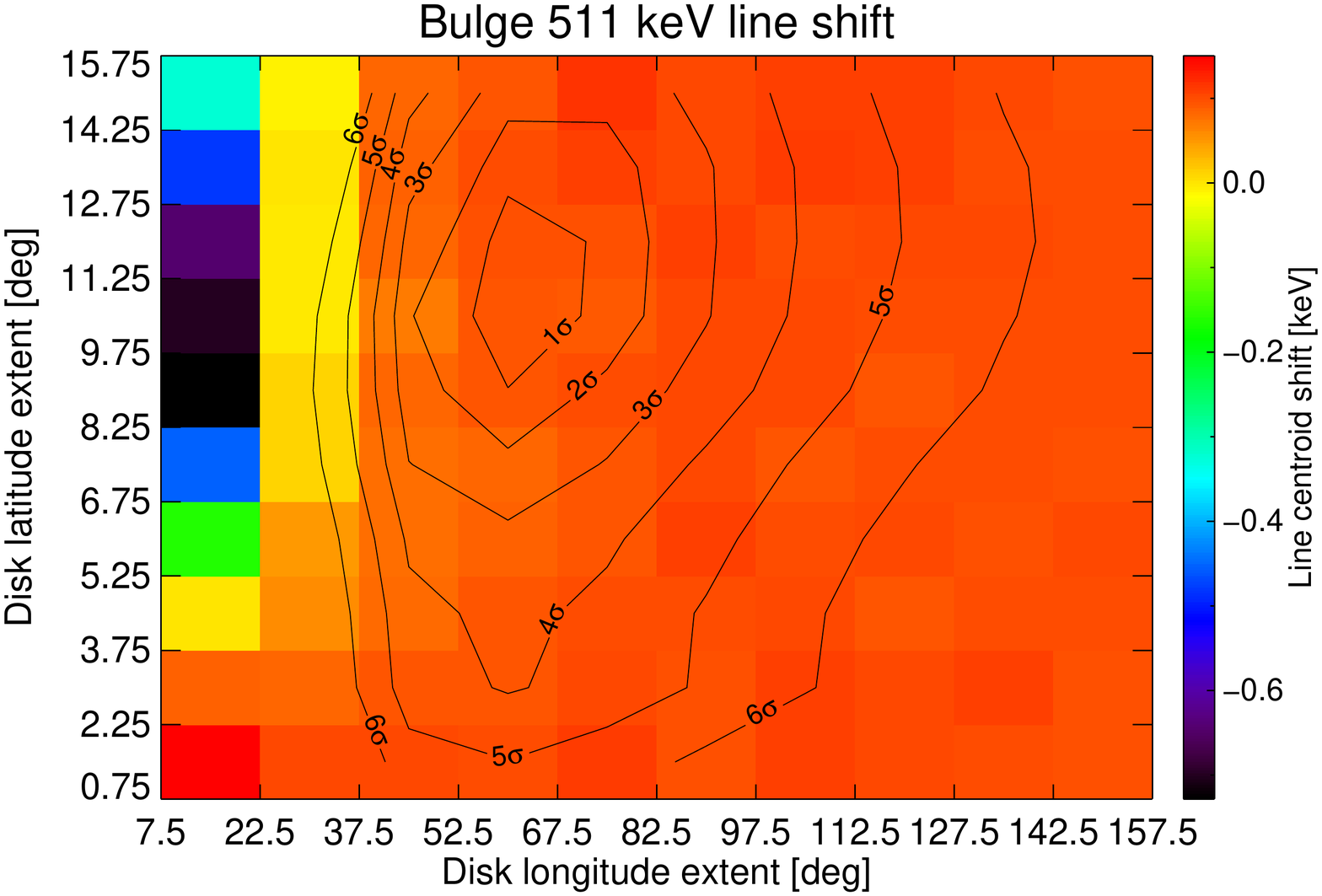}
  \includegraphics[width=0.33\linewidth]{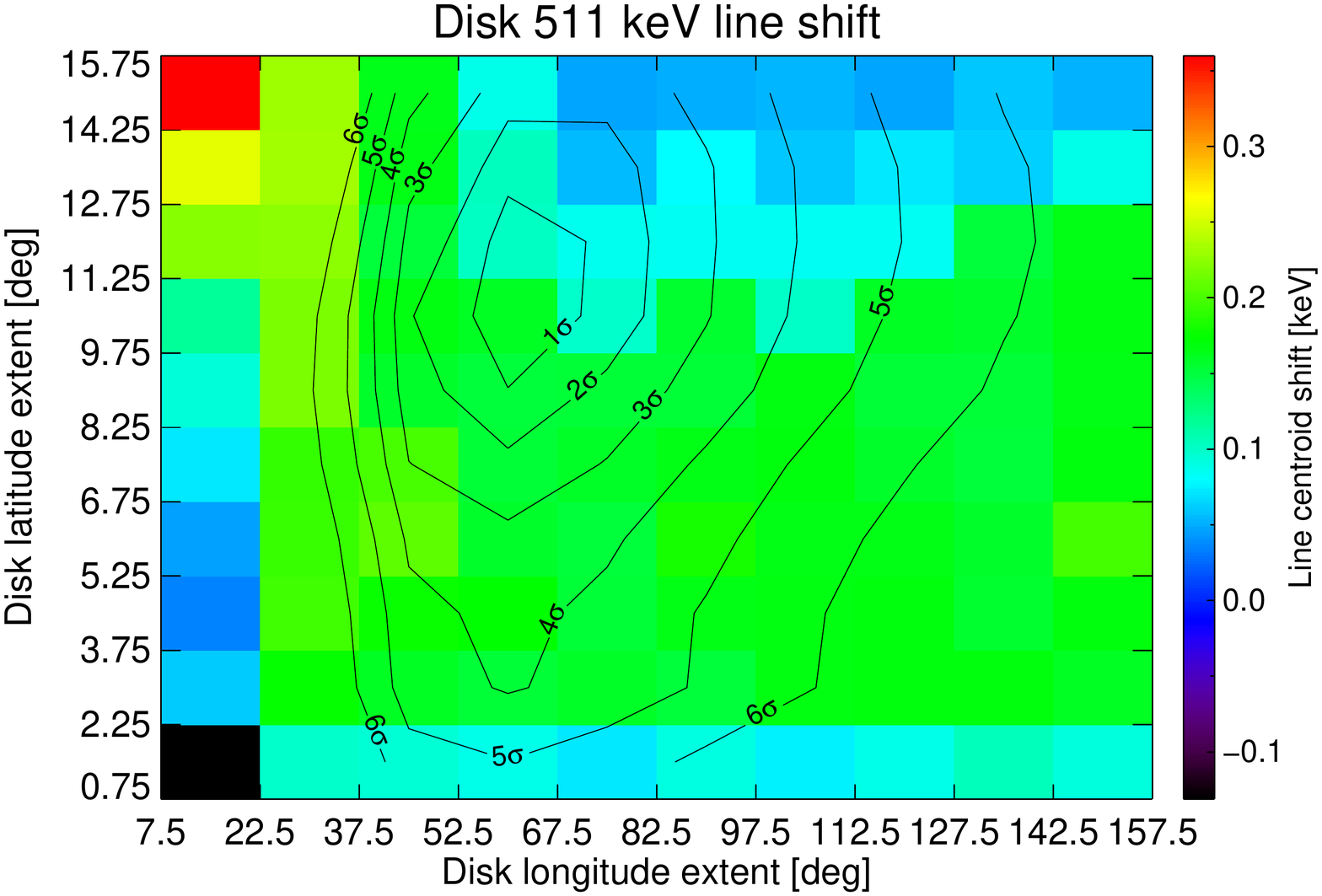}
  \includegraphics[width=0.33\linewidth]{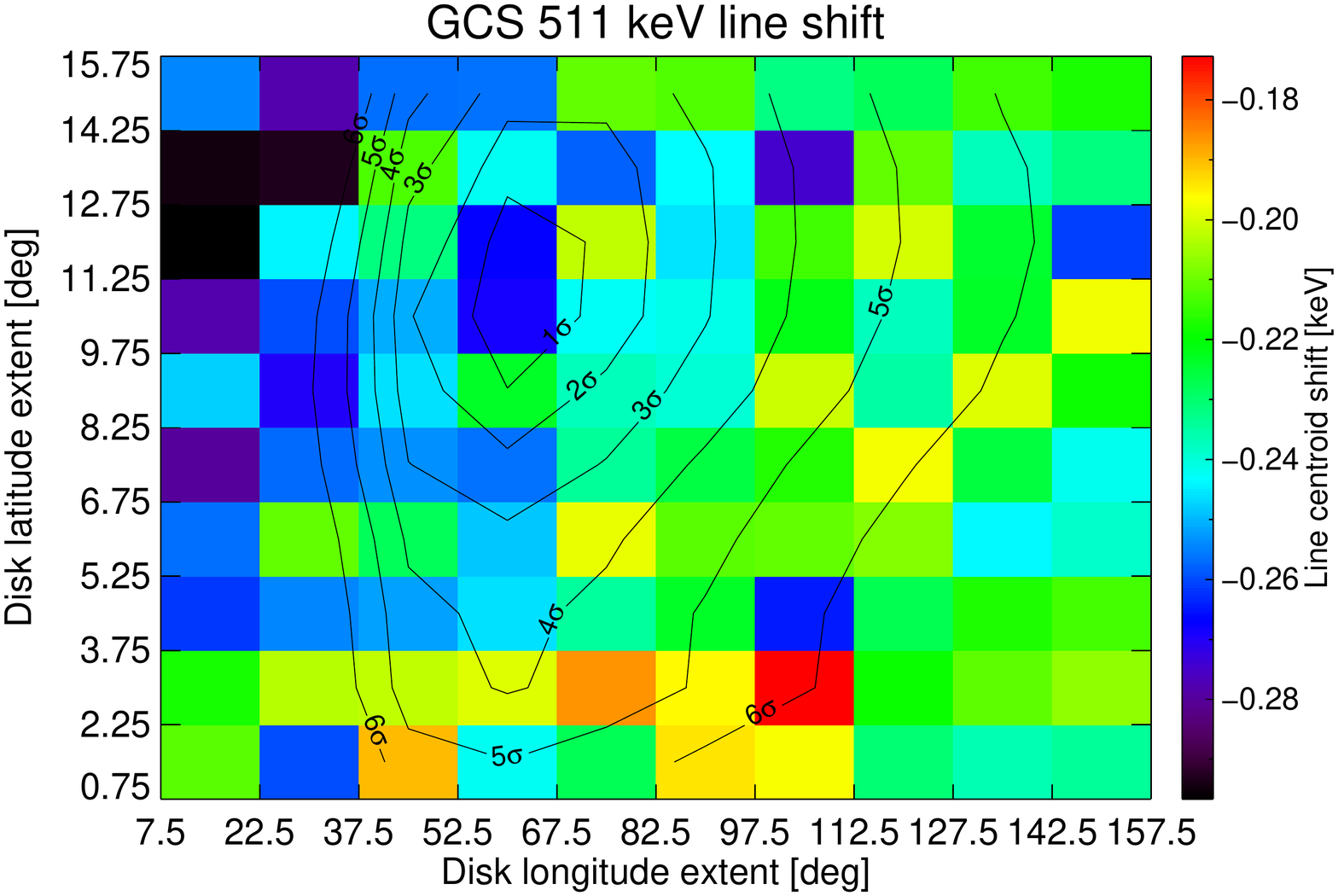}
  \\
  \includegraphics[width=0.33\linewidth]{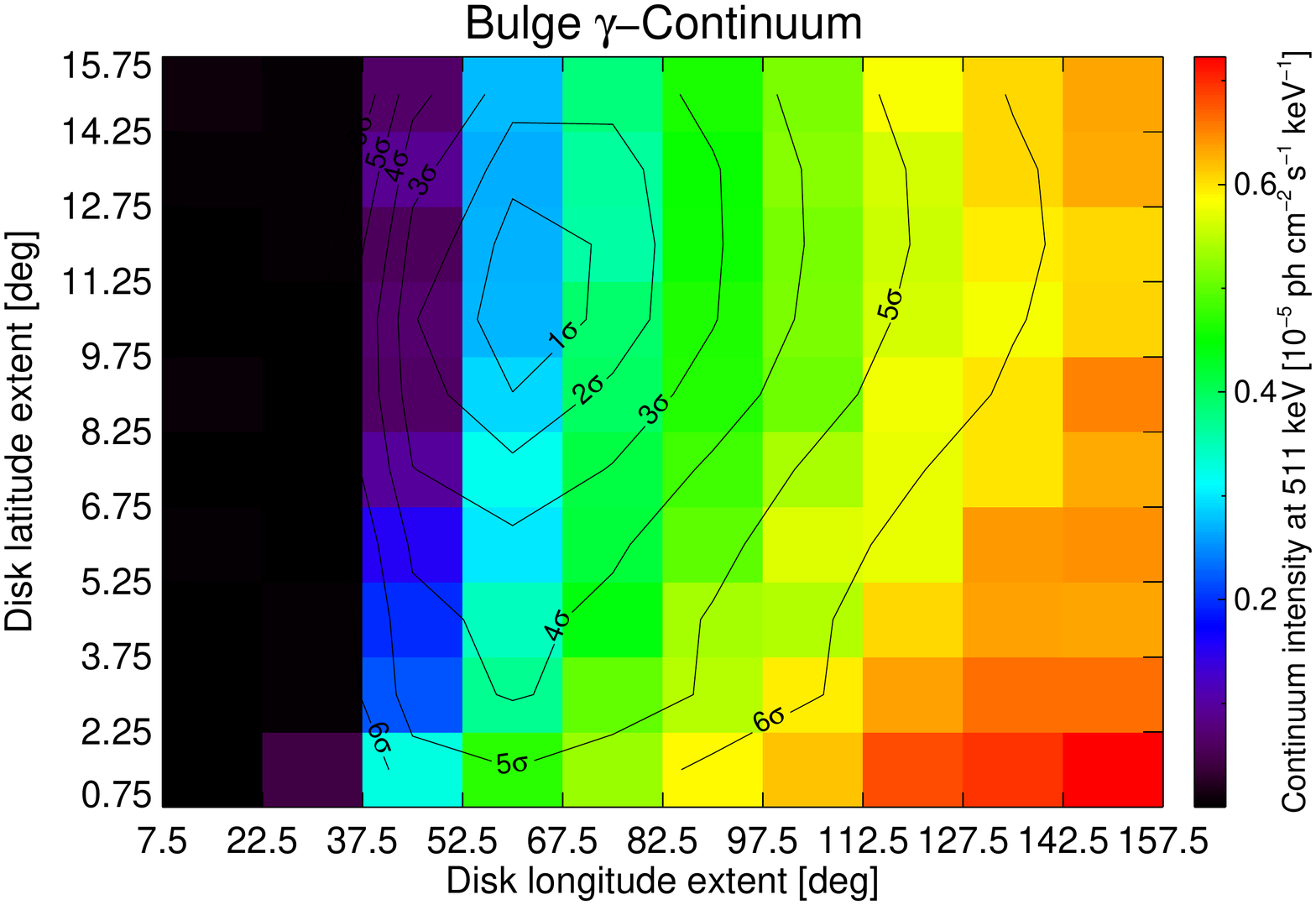}
  \includegraphics[width=0.33\linewidth]{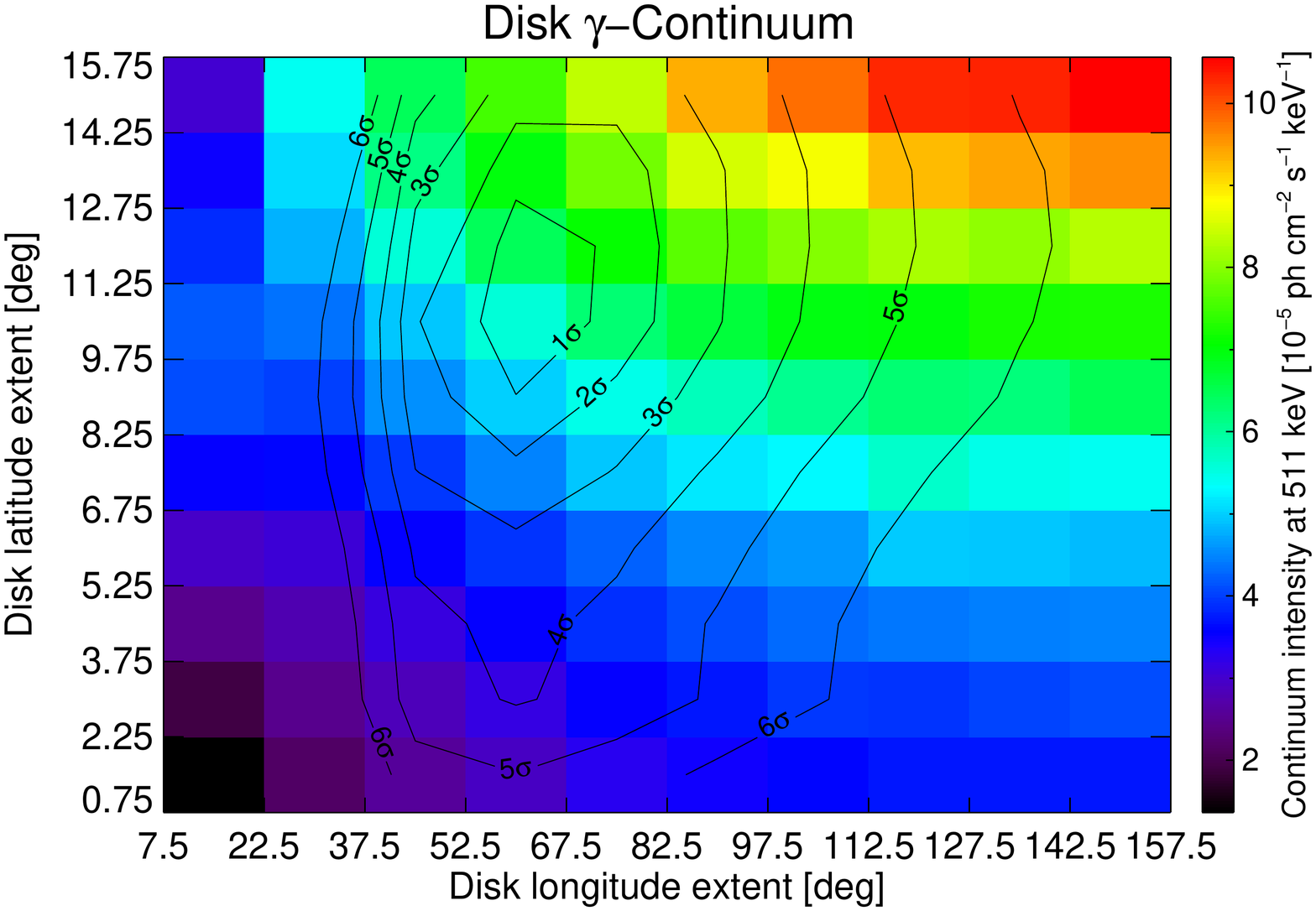}
  \includegraphics[width=0.33\linewidth]{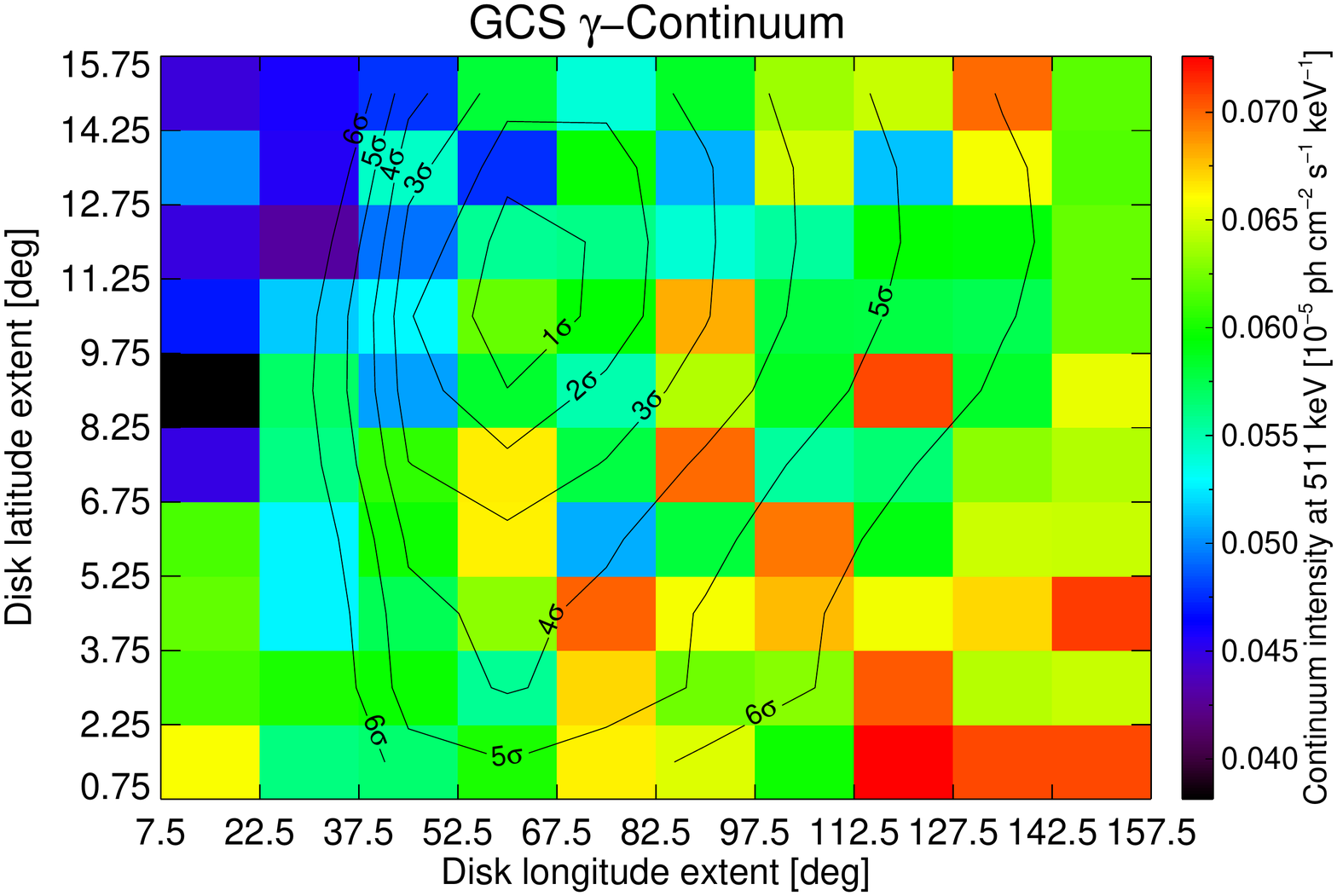}
  \\
  \includegraphics[width=0.33\linewidth]{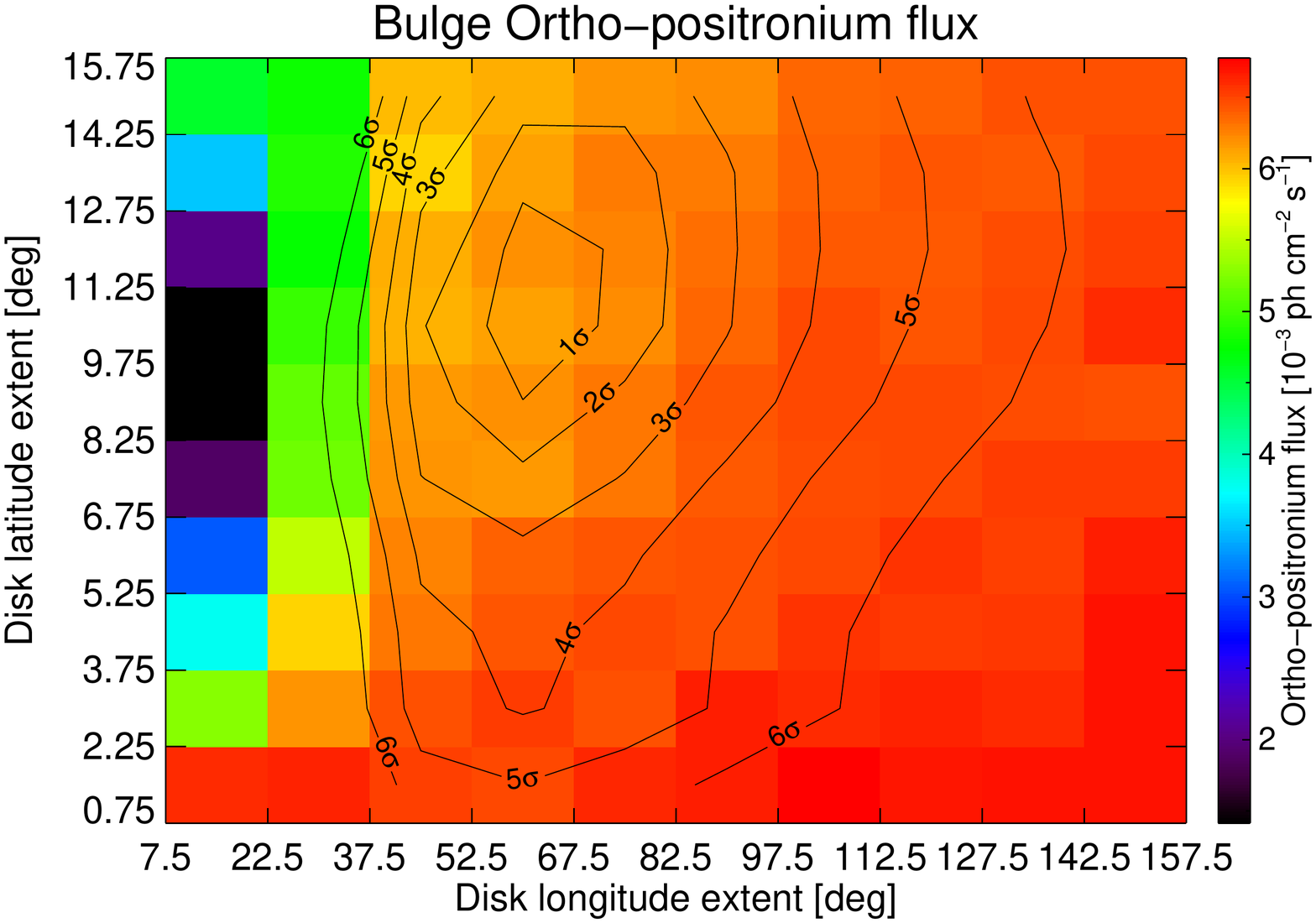}
  \includegraphics[width=0.33\linewidth]{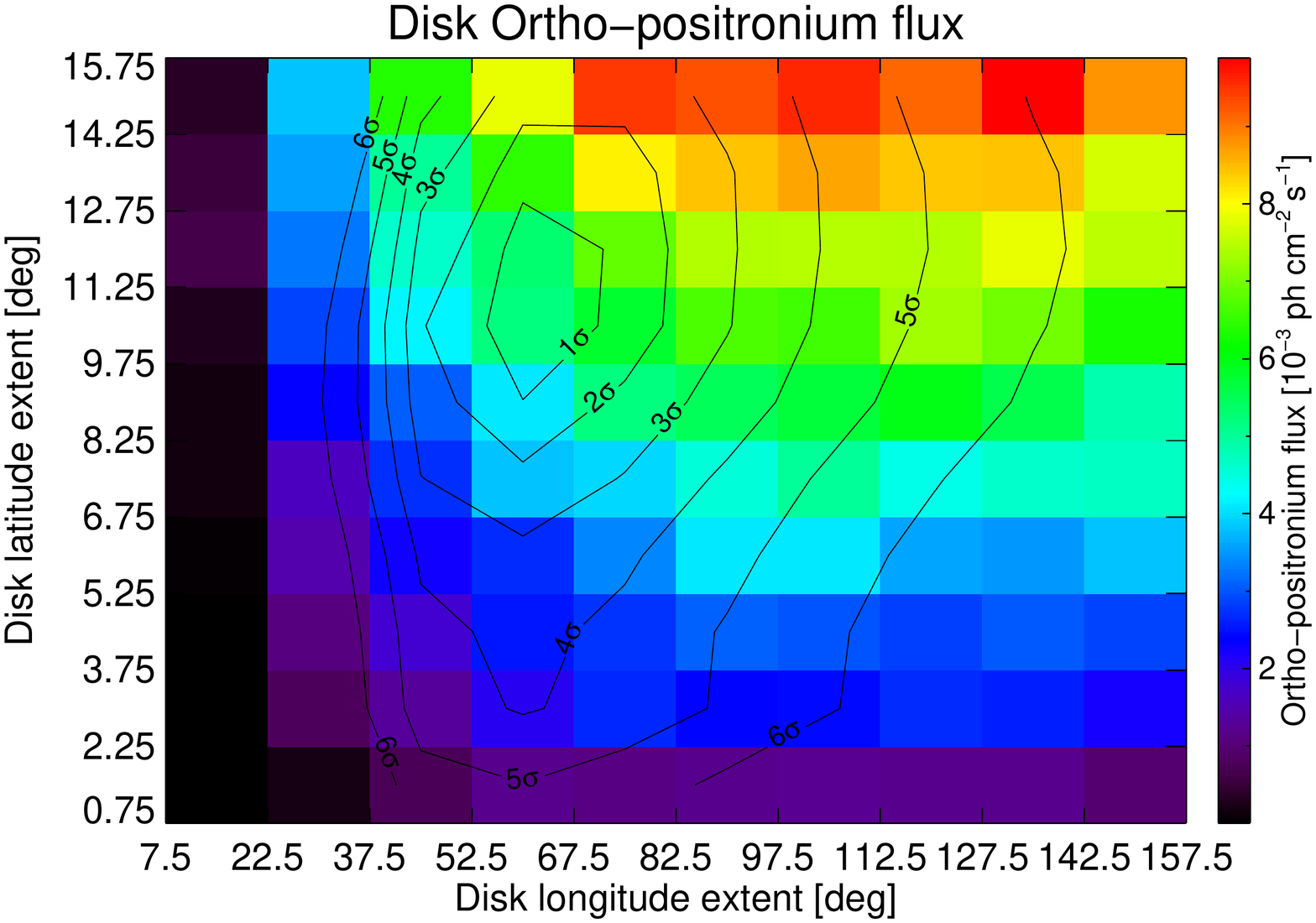}
  \includegraphics[width=0.33\linewidth]{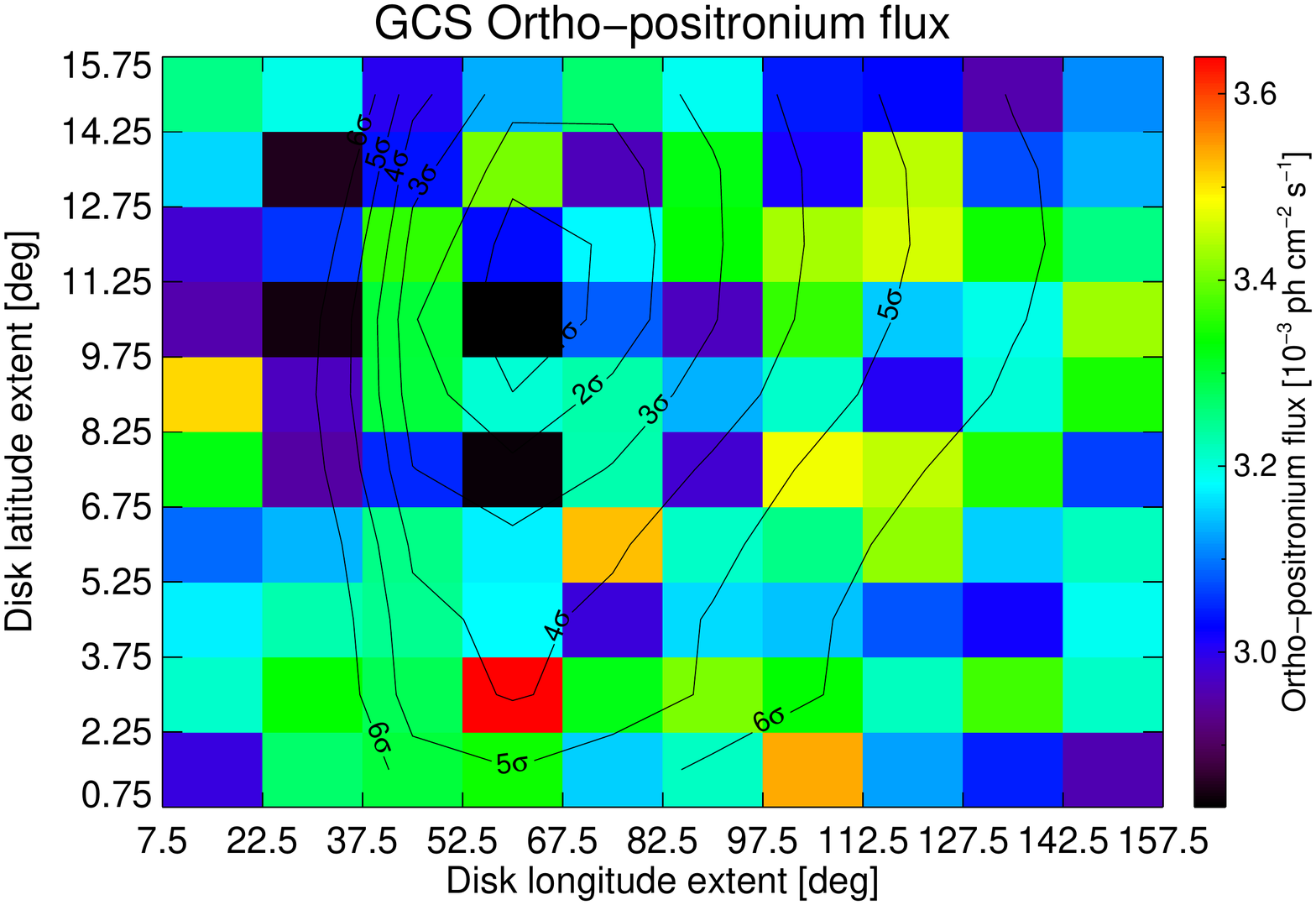}
  \caption{Annihilation line intensity, width, and centroid shift (top $3\times3$ panels), and ortho-positronium and (diffuse) gamma-ray continuum (bottom $3\times2$ panels) of the three main sky components as a function of the selected disk longitude and latitude extent. See figure caption of Fig.~\ref{fig:disk_ext_vs_disk_int} for more details.}
  \label{fig:spec_corr_line}
\end{figure*}

\begin{figure}
  \centering
  \includegraphics[width=\linewidth]{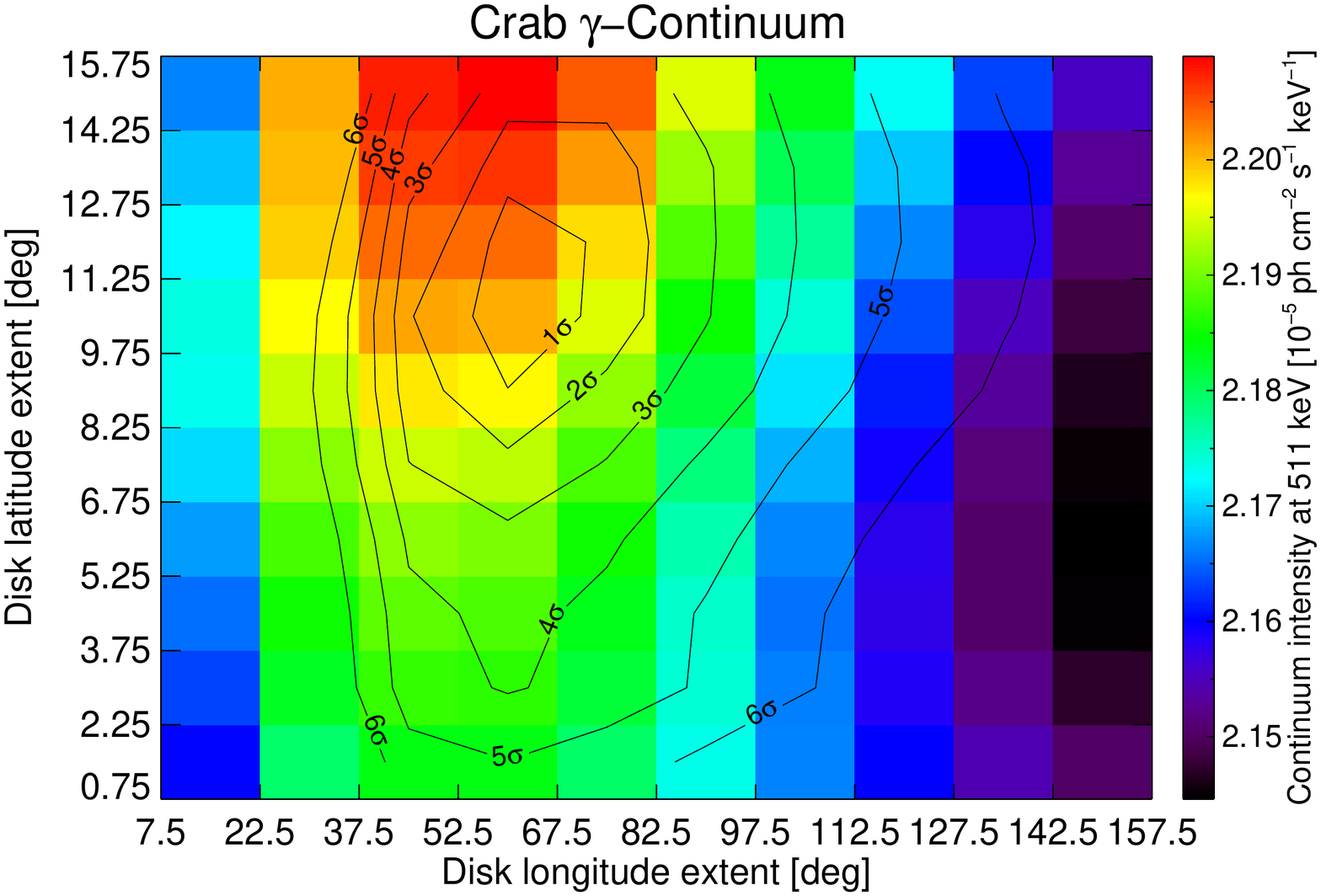} 
   \includegraphics[width=\linewidth]{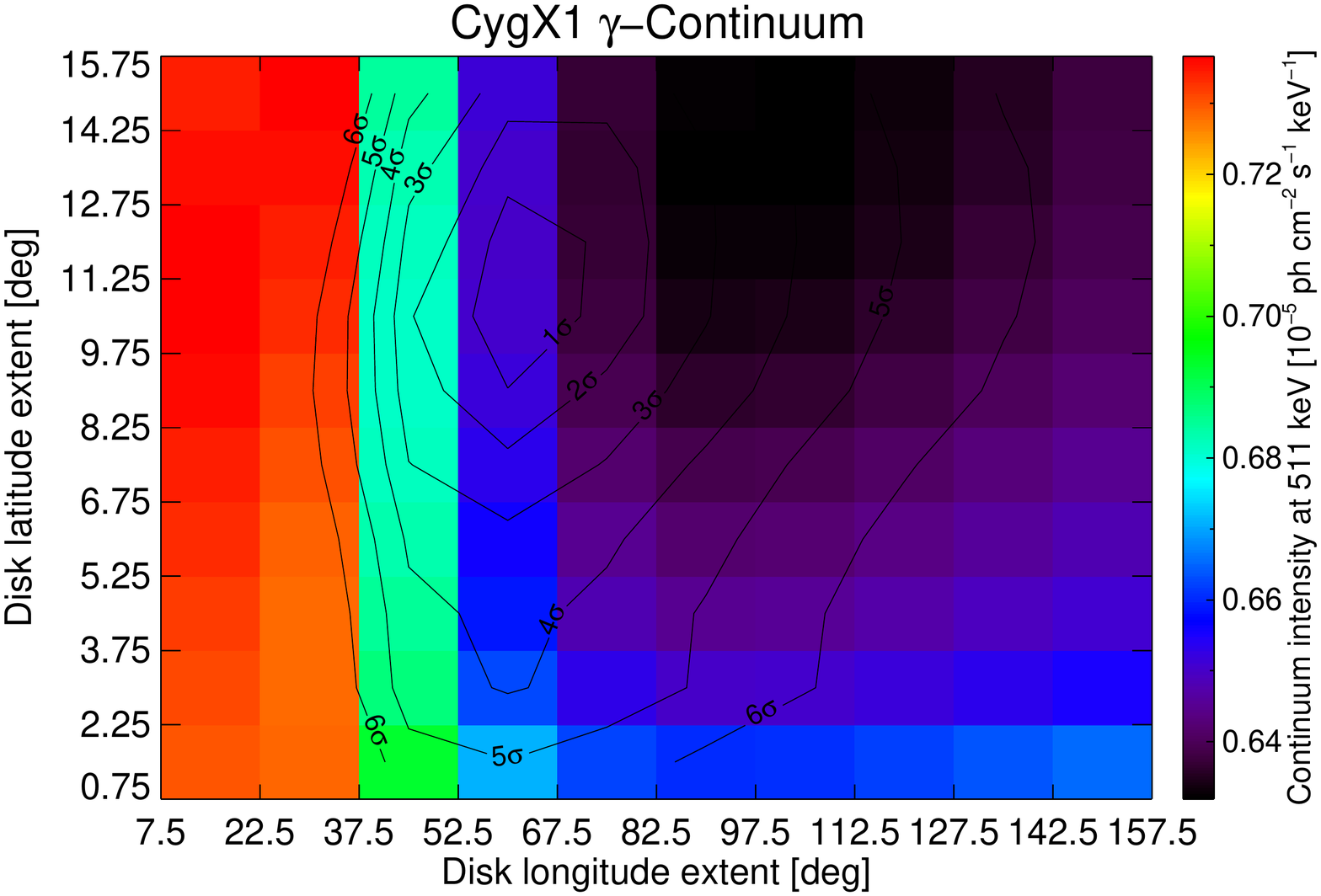} 
 \caption{Continuum flux density of the Crab (top) and Cygnus X-1 (bottom) as a function of the disk size. See figure caption of Fig.~\ref{fig:disk_ext_vs_disk_int} for a detailed description.}
  \label{fig:spec_corr_crabcyg}
\end{figure}

As the bulge morphology is fixed, the longitude and latitude extent of the disk are the most sensitive model parameters to affect the magnitude of cross correlation with the amplitudes of line and continuum, and other spectral parameters in the other sky components. We therefore vary the longitude extent of the disk across a plausible range, and analyse how the fit quality is affected, and how the amplitudes of sky components and their spectral parameters vary with this choice. As shown above for the annihilation line intensity in bulge and disk (Fig.~\ref{fig:disk_ext_vs_disk_int}), the choice for the disk extent around 60$^\circ$ in longitude, and 10.5$^\circ$ in latitude, is preferred. In Fig.~\ref{fig:spec_corr_line}, results of cross correlations between line intensity, line width and line centroid of the three main sky models (Bulge, Disk, GCS) as a function of the disk size (top $3\times3$ panels) are shown. In the bottom $3\times2$ panels of Fig.~\ref{fig:spec_corr_line}, we show the results of cross correlations as they affect the continuum components from ortho-positronium and the large-scale diffuse galactic emission, and the point-source gamma-ray continuum. In Fig.~\ref{fig:spec_corr_crabcyg}, the continuum flux densities of the Crab (top) and Cygnus X-1 (bottom) are illustrated for a varying disk size.

In general, the parameters for the bulge, GCS, the Crab, and Cyg X-1 are not very sensitive to the disk size, except when the disk is modelled with very small size, i.e. $\sigma_l \lesssim 45^\circ$. In this case, the maximum likelihood method approach cannot identify the components separately as they are largely overlapping and the small differences are difficult to disentangle, thus creating confusion. In the case of Cyg X-1 and a short disk, the source is wrongly capturing flux that is probably attributed to the disk. The derived spectral parameters of the disk are the most varying parameters. This results from the chosen shape to model the disk. At increasing disk size in longitude and latitude, the detection of low surface-brightness regions increases, and more line and continuum flux can be found. However, in our 80 bin spectral analysis, the longitude and latitude sizes, albeit biased by other parameters (see Sect.~\ref{sec:bulge_morph_effects}), are constrained very well, as shown by the $1\sigma$ contours in the middle panels of Fig.~\ref{fig:spec_corr_line}, varying by 20\% at most.

As the spectral parameters of the other four sky components are not too sensitive to the disk size, and the disk's spectral parameters are sensitive to the disk size itself, we conclude that the cross correlations are almost negligible with respect to the flux uncertainties in each spectrum, and that the fitted and derived parameters of each component are representative for the component itself.

% ----------------------------------------------------------------------
% ----------------------------------------------------------------------

% ----------------------------------------------------------------------
% ----------------------------------------------------------------------

\subsection{Bulge morphology effects}
\label{sec:bulge_morph_effects}

The focus of our analysis was finding an optimal disk size as derived from a spectral analysis of the 511~keV morphology of the Milky Way, given an already optimised bulge morphology fixed in size (in the following called baseline model). A maximum likelihood fit for the 80 half keV energy bins in the studied energy range of 490 to 530~keV is performed. For each of the 80 bins, 100 tested disk sizes are convolved through the imaging response function (coded mask shadowgram) of the SPI instrument. As the convolution of a large number of disk and bulge models is very computer intense and time-consuming, we refer to \citet{Skinner2014_511} which studied the dependency of the several bulge morphology models in a single 6~keV bin from 508 to 514~keV. When they consider the background model described in this work (see Sect.~\ref{sec:bg_model}), and search for an optimal bulge and disk morphology (modelled by 2D-Gaussians on a bigger grid) simultaneously, a slightly different bulge model is found. The best-fit parameters are listed in Tab.~\ref{tab:alternative_bulge}. As the responses for the disk do not have to be repeated, a similar search for the best-fitting disk size is applied to the alternative bulge morphology (see Tab.~\ref{tab:alternative_bulge}). 

% ----------------------------------------------------------------------
\begin{table}
\caption{Alternative emission model.}              % title of Table
\label{tab:alternative_bulge}      % is used to refer this table in the text
\centering                                      % used for centering table
\begin{tabular}{>{\RaggedRight}p{1.1cm} >{\RaggedLeft}p{1.4cm} >{\RaggedLeft}p{1.4cm} >{\RaggedLeft}p{1.6cm} >{\RaggedLeft}p{1.6cm}}          % centered columns (6 columns)
\hline\hline                        % inserts double horizontal lines
%Name & Galactic Longitude Position [deg] & Galactic Latitude Position [deg] & Longitude extent (FWHM) [deg] & Latitude extent (FWHM) [deg] \\
Component & G. Lon. position [deg] & G. Lat. position [deg] & Lon. extent (FWHM) [deg] & Lat. extent (FWHM) [deg] \\
\hline                                   % inserts single horizontal line
    NB                        & $-0.75$   & $0.00$ & $9.04$   & $7.40$  \\      % inserting body of the table
    BB                        & $0.00$    & $0.00$  & $49.31$  & $24.65$ \\
    Disk\tablefootmark{a}     & $0.00$    & $0.00$  & $211.93$ & $28.26$ \\
\hline                                             %inserts single line
\end{tabular}
\tablefoot{
\tablefoottext{a}{The disk extent has been chosen according to a 2D grid scan for a total maximum likelihood over all 80 bins.}
}
\end{table}
% ----------------------------------------------------------------------

An image of this morphology is almost indistinguishable, compared to Fig.~\ref{fig:SPI-imagemodel}. The NB and BB components are now even more extended: Both components are now rather elongated in longitude. The latitude FWHM of the BB is ranging up to $\approx55^\circ$ and can therefore be interpreted as a halo-like morphology. We use this alternative model of the central part of the galaxy as a robustness check of our result. As the bulge is becoming larger, also its flux is increased ($(1.51\pm0.23)\times10^{-3}\mathrm{~ph~cm^{-2}~s^{-1}}$), although the celestial line width appears narrower ($(2.35\pm0.34)~\mathrm{keV}$). The values are consistent with the baseline model within $2\sigma$. As the overlap between bulge (both NB and BB) and disk is also bigger, the bulge is capturing more of the diffuse galactic continuum emission ($(2.13\pm0.91)\times10^{-5}\mathrm{~ph~cm^{-2}~s^{-1}~keV^{-1}}$). The disk size is increased by about 15\%, now covering almost the whole latitude range with low surface-brightness regions, and the flux is increased by $\sim11\%$. Within uncertainties, the spectral parameters are still consistent. In Tab.~\ref{tab:alternative_pars}, the fitted spectral parameters of the alternative bulge morphology are listed. The parameters agree when compared to the baseline model. Especially the positron annihilation conditions in the ISM of the respective components are unchanged when changing the emission model which is encouraging and validates the findings.

% ----------------------------------------------------------------------
\begin{table*}
\caption{Spectral parameters for each component from the alternative bulge emission model. Continuum flux densities are given as the value at 511~keV in units of $10^{-5}~\mathrm{ph~cm^{-2}~s^{-1}~keV^{-1}}$, line and ortho-positronium fluxes are given in $10^{-4}~\mathrm{ph~cm^{-2}~s^{-1}}$, FWHM of the celestial emission line in keV, and the centroid shift $\Delta E_{0} = E_{peak} - E_{lab}$ in keV.}              % title of Table
\label{tab:alternative_pars}      % is used to refer this table in the text
\centering                                      % used for centering table
%\begin{tabular}{c | r r r r r r r r r}          % centered columns (6 columns)
\begin{tabular}{l | >{\RaggedLeft}p{1.45cm} >{\RaggedLeft}p{1.45cm} >{\RaggedLeft}p{1.45cm} >{\RaggedLeft}p{1.45cm} >{\RaggedLeft}p{1.45cm} >{\RaggedLeft}p{1.45cm} >{\RaggedLeft}p{1.45cm}}
\hline\hline                        % inserts double horizontal lines
%Name & Galactic Longitude Position [deg] & Galactic Latitude Position [deg] & Longitude extent (FWHM) [deg] & Latitude extent (FWHM) [deg] \\
                           & Cont. flux dens. & Line Flux   & FWHM        &  $\Delta E_0$   &  o-Ps Flux   & Pos. frac. & $\chi^2$/d.o.f. \\
\hline                                   % inserts single horizontal line
    The Bulge                           & $2.31(91)$    & $15.1(2.3)$ & $2.35(34)$ & $ 0.24(17)$  & $82.6(31.6)$  & $1.05(9)$   & $61.53/74$ \\  
    Disk ($-180^\circ < l < 180^\circ$) & $5.76(1.69)$  & $18.4(6.0)$ & $3.25(98)$ & $-0.08(36)$  & $41.8(48.9)$  & $0.80(39)$  & $69.48/74$ \\
    GCS                                 & $0.09(5)$     & $0.9(2)$    & $3.04(54)$ & $-0.24(28)$  & $2.3(1.7)$    & $0.85(23)$  & $63.98/74$ \\
    Crab\tablefootmark{a}               & $2.20(7)$     & $<0.7$      & $-(-)$     & $-(-)$       & $-(-)$        & $-(-)$      & $66.88/78$ \\
    Cygnus X-1\tablefootmark{a}         & $0.66(6)$     & $<0.2$      & $-(-)$     & $-(-)$       & $-(-)$        & $-(-)$      & $73.38/78$ \\
    \hline
    Galaxy (total)                      & $10.71(1.29)$ & $34.3(3.9)$ & $2.77(29)$ & $ 0.10(11)$  & $137.9(44.3)$ & $0.97(9)$   & $73.70/74$ \\
\hline                                             %inserts single line
\end{tabular}
\tablefoot{
\tablefoottext{a}{For the Crab and Cyg X-1, no annihilation line has been detected, and $2\sigma$ upper limits are given for the flux.}
}
\end{table*}
% ----------------------------------------------------------------------

This alternative description of the bulge yields a total maximum (log)-likelihood that is larger by a value of $6.86$ (summed over all 80 bins), compared to our used model. In particular, the improvement in the fit is randomly distributed\footnote{KS-test probability: 8.2\%; minimum: $-1.63$; maximum: $2.46$.} among the spectral energy range of interest with a mean of $0.09$, and a variance of $0.55$. Although the test-statistics is better for the alternative model, we prefer the baseline model.

In the baseline model, the reduced overlap between bulge and disk improves the separability of those two components and allows for a distinct discussion of their spectral details. An increased overlap also increases the intensity correlations and therefore the uncertainties on each spectral data point. This is reflected in the fitted parameters of the alternative model, compared to the baseline model.

The signal-to-noise ratios, summed over all 80 energy bins,
\begin{equation}
\sqrt{\sum_{i=1}^{80}\left(\frac{\mu_i}{\sigma_i}\right)^2}\mathrm{,}
\label{eq:signal-to-noise}
\end{equation}
for each component are listed in Tab.~\ref{tab:stnratio}, comparing the signal strengths for both tested morphologies. In Eq.~(\ref{eq:signal-to-noise}), $\mu_i$ is the values of the data point for an energy bin $i$ in a particular spectrum, and $\sigma_i$ the respective uncertainty on this data point.

% ----------------------------------------------------------------------
\begin{table}
\caption{Total signal-to-noise ratio, over the 80 analysed energy bins for each celestial model component in one coherent maximum likelihood analysis.}              % title of Table
\label{tab:stnratio}      % is used to refer this table in the text
\centering                                      % used for centering table
\begin{tabular}{>{\RaggedRight}p{1.6cm} | >{\RaggedLeft}p{0.8cm} >{\RaggedLeft}p{0.8cm} >{\RaggedLeft}p{0.8cm} >{\RaggedLeft}p{0.8cm} >{\RaggedLeft}p{0.8cm}}          % centered columns (6 columns)
\hline\hline                        % inserts double horizontal lines
%Name & Galactic Longitude Position [deg] & Galactic Latitude Position [deg] & Longitude extent (FWHM) [deg] & Latitude extent (FWHM) [deg] \\
    Morphology              & Bulge     & Disk   &   GCS    & Crab     & Cyg X-1 \\
\hline                                   % inserts single horizontal line
    Baseline                & $29.0$   & $17.3$ & $10.8$   & $26.9$   & $13.5$      \\      % inserting body of the table
    Alternative             & $15.6$   & $12.2$ & $11.7$   & $26.9$   & $13.5$      \\
\hline                                             %inserts single line
\end{tabular}

\end{table}
% ----------------------------------------------------------------------

The signal-to-noise ratio for the two continuum sources, the Crab and Cygnus X-1, is not affected by the choice of the diffuse emission morphology. The uncertainties on the diffuse emission components of the bulge and the disk are very sensitive to the chosen shape and extents of the morphology because very low surface brightness components, as are apparent in the alternative model ($|l|<30^{\circ}$, $|b|>30^{\circ}$), automatically introduce larger uncertainties on these components which reflects in a statistically better fit.

Both models provide statistically acceptable fits to the data and are consistent with each other within $2\sigma$ uncertainties. Especially, the disk latitude extent is found rather large for both models.

\subsection{Comparing individual spectra}
\label{sec:comp_spec}
Our spectra at 0.5~keV binning in the 490-530~keV band around the 511 keV annihilation line allow a new look at Galactic positron annihilation, as separation among different components of bulge, disk, and central source has been achieved. Spectral result parameters are listed in Tab.~\ref{tab:spec_deriv_params}. We investigate differences of the derived spectra in their entireties, by either comparing each spectrum with each other or to the total annihilation spectrum of the Milky Way. We are only comparing the annihilation part of the spectra, i.e. by subtracting the best-fit power-law shaped gamma-continuum contribution, $C(E)$. Here we ask how the best-fit spectrum of one component may be acceptable as a representation of data from another component. In detail, we fix the spectral model with the best-fit parameters ($I_L$, $I_O$, $E_0$, $\sigma$) of one spectrum and fit this template spectrum with only one global amplitude to all other spectra, separately. The difference in the d.o.f. is consequently 3. In Tab.~\ref{tab:chi2_test_spec}, we show the significance in units of $\sigma$ as calculated by a $\chi^2$-test with 3 d.o.f.. The table reads as follows: Line by line, the spectra (data points per energy bin) of the sky components are fitted by the templated spectrum (smooth function) of the other spectra (columns). 

% ----------------------------------------------------------------------
\begin{table}[!ht]
\caption{Spectral model comparisons of the celestial components, GCS, Bulge, eastern and western hemisphere of the Disk, and the total/combined Milky Way (incorporating all correlations among different components, i.e. a mixture of all models). The probability values are given in units of $\sigma$, reflecting the tension of one templated best-fit spectrum being representative for another spectrum, or the combined Milky Way spectrum.}              % title of Table
\label{tab:chi2_test_spec}      % is used to refer this table in the text 
\centering                                      % used for centering table
%\begin{tabular}{c | r r r r r r r r r}          % centered columns (6 columns)
\begin{tabular}{l | >{\RaggedLeft}p{0.8cm} >{\RaggedLeft}p{0.8cm} >{\RaggedLeft}p{1.0cm} >{\RaggedLeft}p{1.0cm} | >{\RaggedLeft}p{0.8cm} }
\hline\hline                        % inserts double horizontal lines
%Name & Galactic Longitude Position [deg] & Galactic Latitude Position [deg] & Longitude extent (FWHM) [deg] & Latitude extent (FWHM) [deg] \\
    Sig. $[\sigma]$     & GCS      & Bulge  &   Disk ($l>0$) &  Disk ($l<0$)   &  Galaxy  \\
\hline                                   % inserts single horizontal line
    GCS                         & $0.0$    & $0.6$  & $1.0$          & $1.5$           & $0.8$    \\  
    Bulge                       & $4.9$    & $0.0$  & $5.1$          & $6.1$           & $2.9$    \\
    Disk ($l>0$)                & $1.8$    & $1.0$  & $0.0$          & $1.1$           & $0.2$    \\
    Disk ($l<0$)                & $2.0$    & $1.4$  & $1.1$          & $0.0$           & $0.5$    \\
    \hline
    Galaxy                      & $3.5$    & $1.6$  & $1.2$          & $2.0$           & $0.0$   \\
\hline                                              %inserts single line
\end{tabular}
\end{table}
% ----------------------------------------------------------------------

The spectrum for the GCS can easily be represented by the spectral shapes of the bulge ($0.6\sigma$) and the two disk halves ($1.0\sigma$, $1.5\sigma$), whereas the opposite is statistically more discouraged ($4.9\sigma$ bulge, $1.8\sigma$ and $2.0\sigma$ disk). The spectrum of the bulge cannot be represented by the spectrum of the disk ($5.1\sigma$, $6.1\sigma$), conversely, the disk spectrum can be represented by the bulge spectrum ($1.0\sigma$ and $1.4\sigma$). The two disk halves as a whole differ by $1.1\sigma$. The spectra, however, only differ in the widths of the spectral line and not in the centroid or o-Ps flux, so that the significance is actually higher ($2.8\sigma$). The Galaxy as a whole cannot be represented by only the best-fit spectrum of the GCS ($3.5\sigma$). For the bulge, and the two disk halves, the tension is not as strong ($1.6\sigma$ bulge, $1.2\sigma$ and $2.0\sigma$ disk). However, all individual spectral components can be represented by the mixed Milky Way template spectrum, except for the bulge in which a $2.9\sigma$ discrepancy is found.

\end{appendix}
% ----------------------------------------------------------------------
\end{document}